\documentclass[fleqn,usenatbib]{mnras}

\usepackage[T1]{fontenc}

\usepackage{graphicx}	
\usepackage{amsmath}	
\usepackage{amssymb}	

\usepackage{threeparttable}

\providecommand{\sorthelp}[1]{}

\title[Dust-to-neutral gas ratio of the IVCs and HVCs]{Dust-to-neutral gas ratio of the intermediate and high velocity \ion{H}{i} clouds derived based on the sub-mm dust emission for the whole sky}

\author[Hayakawa and Fukui]{
Takahiro Hayakawa,$^{1}$\thanks{E-mail: t.hayakawa@a.phys.nagoya-u.ac.jp (TH)}
and Yasuo Fukui$^{1}$
\\
$^{1}$Department of Physics, Nagoya University, Furo-cho Chikusa-ku, Nagoya 464-8602, Japan
}

\date{Accepted XXX. Received YYY; in original form ZZZ}

\pubyear{2024}

\usepackage{newtxtext,newtxmath}

\begin{document}
\label{firstpage}
\pagerange{\pageref{firstpage}--\pageref{lastpage}}
\maketitle

\begin{abstract}
We derived the dust-to-\ion{H}{i} ratio of the intermediate-velocity clouds (IVCs), the high-velocity clouds (HVCs), and the local \ion{H}{i} gas, by carrying out a multiple-regression analysis of the 21\,cm \ion{H}{i} emission combined with the sub-mm dust optical depth.
The method covers over 80 per cent of the sky contiguously at a resolution of 47\,arcmin and is distinguished from the absorption line measurements toward bright galaxies and stars covering a tiny fraction of the sky.
Major results include that the ratio of the IVCs is in a range of 0.1--1.5 with a mode at 0.6 (relative to the solar-neighbourhood value, likewise below) and that a significant fraction, $\sim 20$ per cent, of the IVCs include dust-poor gas with a ratio of $<0.5$.
It is confirmed that 50 per cent of the HVC Complex~C has a ratio of $<0.3$, and that the Magellanic Stream has the lowest ratio with a mode at $\sim 0.1$.
The results prove that some IVCs have low metallicity gas, contrary to the previous absorption line measurements.
Considering that the recent works show that the IVCs are interacting and exchanging momentum with the high-metallicity Galactic halo gas, we argue that the high-metallicity gas contaminates a significant fraction of the IVCs.
Accordingly, we argue that the IVCs include a significant fraction of the low metallicity gas supplied from outside the Galaxy as an alternative to the Galactic-fountain model.
\end{abstract}




\defcitealias{2014ApJ...796...59F}{F14}
\defcitealias{2015ApJ...798....6F}{F15}
\defcitealias{2017PASJ...69L...5F}{F17}
\defcitealias{2017ApJ...838..132O}{O17}
\defcitealias{2019ApJ...878..131H}{H19}
\defcitealias{2019ApJ...871...44T}{T19}
\defcitealias{2021PASJ...73S.117F}{F21}

\section{Introduction}
\label{sec:introduction}

The high-velocity clouds (HVCs) and their slower-moving counterparts, the intermediate-velocity clouds (IVCs), are \ion{H}{i} clouds with large velocities that a simple model of the Galactic rotation can not explain.
Typically observed at positions distant from the Galactic plane and predominantly in a negative radial velocity range.
The early discoveries were made over a half-century ago (e.g., \citealt[][]{1952PASP...64..312M,1961ApJ...133...11M,1963CRAS..257.1661M,1963BAN....17..203S,1966BAN....18..405B}; see a review by \citealt{2004ASSL..312....1W} for research history up to 1999).
There have been several efforts to constrain the distance to the IVCs/HVCs using the absorption-line bracketing technique.
The IVCs are located relatively close, and have typical $z$ heights of $\sim 1$--2\,kpc (e.g., \citealt{2008ApJ...672..298W,2022MNRAS.513.3228L}, see also compilations by \citealt{2001ApJS..136..463W,2004ASSL..312..195V} and references therein), and are likely to be the disc-halo interface objects, while the major HVCs are further away, several to $\sim 10$\,kpc above the disc \citep[e.g.,][]{1999Natur.400..138V,2006ApJ...638L..97T,2008ApJ...684..364T,2007ApJ...670L.113W,2008ApJ...672..298W,2011MNRAS.415.1105S,2015A&A...584L...6R,2022MNRAS.513.3228L} and belong to the inner halo.
One of the important questions is whether the IVCs and HVCs are different halo-cloud populations (i.e., gas comes from the disc, and the one fuels the disc) or just the same population of objects with different heights and radial velocities.

The metallicity is a crucial parameter for pursuing the origin and has traditionally been measured by observing absorption lines. 
The previous works presented that the prominent HVCs in the Complexes A and C have sub-solar metallicities $Z/Z_{\sun}\sim 0.1$--$0.3$ (\citealt{2001ApJ...559..318R,2003AJ....125.3122T,2003ApJ...585..336C,2007ApJ...657..271C,2004ApJS..150..387S,2023ApJ...946L..48F}, see also compilations by \citealt{2001ApJS..136..463W}) and are of extra-galactic origin.
The IVCs have been reported to have near-solar metallicity of $\sim 0.5$--1.0 solar (\citealt{2001ApJS..136..463W} and references therein; \citealt[][]{2001ApJ...549..281R,2001ApJ...559..318R}; \citealt{2004ApJS..150..387S}) and are usually explained by the so-called Galactic-fountain model in which hot gas ejected from the disc by the stellar feedback falls back in the form of neutral clouds \citep{1976ApJ...205..762S,1980ApJ...236..577B}.
However, the measured metallicities often have rather large uncertainties due to such factors as ionization states and interstellar depletion (\citealt{2004ASSL..312..195V} summarised the possible problems in the metallicity determination in their section~5).
In addition, these measurements are made at a limited number of locations because of a tiny number of bright background objects and do not lead to firmly establish the metallicity of the IVC.

An alternative method for measuring metallicity is to use dust emissions.
It is a reasonable proxy for absorption measurements because the dust consists of a significant fraction of heavy elements, perhaps comparable to that in the gas phase (see also discussion in Section \ref{subsec:present_vs_absorption}).
For this purpose, the 100\,$\micron$ emission obtained with \textit{IRAS} is often used.
However, it is not a valid measure of the interstellar dust mass.
In the Rayleigh-Jeans regime, all components having various temperatures in a line-of-sight contribute to the emission proportionally to their masses, whereas lower-temperature components contribute little in the Wien regime.
Generally, only in the Rayleigh-Jeans regime can we measure all dust mass.
In interstellar dust, the peak wavelength of the modified Planck function is around 100\,$\micron$, and the sub-mm wavelength satisfies the Rayleigh-Jeans regime.
The Planck Collaboration obtained such sensitive sub-mm dust emission over the whole sky at a 5\,arcmin resolution and derived dust optical depth at 353\,GHz (850\,$\micron$), $\tau_{353}$, by fitting the modified Planck function at four wavelengths from 100 to 850\,$\micron$ obtained by \textit{Planck} and \textit{IRAS} \citep{planck2013-p06b,planck2016-XLVIII}.

Previous works by \citet[F14 and F15 hereafter]{2014ApJ...796...59F,2015ApJ...798....6F}, \citet[O17]{2017ApJ...838..132O}, \citet[H19]{2019ApJ...878..131H}, and \citet{2019ApJ...884..130H} showed that $\tau_{353}$ characterises well the linearity of the dust emission in a number of regions of the Milky Way.
The scheme was also applied to the Large Magellanic Cloud (LMC).
As such, \citet[F17]{2017PASJ...69L...5F} used the linear relationship between the velocity-integrated intensity of \ion{H}{i} line, $W_{\ion{H}{i}}$, and $\tau_{353}$, and estimated that the \ion{H}{i} ridge including R136 has a factor of two lower dust-to-\ion{H}{i} ratio than the optical stellar Bar region in LMC.
In addition, \citet[T19]{2019ApJ...871...44T} presented that the N44 region near the \ion{H}{i} Ridge in the LMC has a $\sim 30$ per cent lower dust-to-\ion{H}{i} ratio than that of the Bar region.
Most recently, \citet[F21 hereafter]{2021PASJ...73S.117F} applied the method to IVC~86$-$36 in the Pegasus-Pisces (PP) Arch and derived a dust-to-\ion{H}{i} ratio relative to the low-velocity (LV) ISM (considered to be in the solar neighbourhood) of $\sim 0.2$, strongly suggesting that PP~Arch originated in a dust-poor environment but not in the disc.
Their measurements of the dust-to-gas ratio significantly improved the preceding metallicity measurements by optical/ultraviolet atomic absorption lines of the background stellar spectrum in the IVC \citep{1997ApJ...475..623F}.
Although the absorption-line results suggest the subsolar metal abundance, the abundance values vary by $\sim 0.5$\,dex depending on the atomic species, leaving the metallicity unquantified. 
Considering the whole above, we judge the dust-to-\ion{H}{i} ratio measurements \citepalias{2017PASJ...69L...5F,2019ApJ...871...44T,2021PASJ...73S.117F} to be most appropriate and adopt it in the present work with an aim to extend the metallicity measurement of the \ion{H}{i} gas over the whole sky outside the Galactic plane.

The aim of the present paper is to use $\tau_{353}$ and the 21\,cm \ion{H}{i} emission and to derive the dust-to-\ion{H}{i} ratio in the IVCs and possibly the other \ion{H}{i} components.
It is not yet examined well if the low column density \ion{H}{i} gas in the IVCs and HVCs have similar dust properties to the Galactic disc, and the extension of the dust emission to the halo \ion{H}{i} gas is a challenge.
Nonetheless, we believe that the paper benefits the community by reporting possible new aspects of the dust emission in these \ion{H}{i} gases as well as the evolutionary trend of the Galactic ISM.
The \ion{H}{i} is an important mass reservoir, while the ionized gas may carry a large mass similar to the neutral \ion{H}{i} gas.
We focus on the neutral \ion{H}{i} gas in the present paper because the dust-to-gas ratio in \ion{H}{i} is one of the essential factors giving information on the chemical evolution.
The paper is organized as follows: Section~\ref{sec:datasets} describes the datasets used, and Section~\ref{sec:properties} gives the dust and gas properties analysed by the present work.
Section~\ref{sec:analyses} shows the dust-to-\ion{H}{i} ratio distribution in middle/high latitudes including the individual outstanding components.
Section~\ref{sec:discussion} gives a discussion focusing on the implications of the IVC dust-to-\ion{H}{i} ratio.
Section~\ref{sec:conclusions} concludes the present work.

\begin{figure*}
\includegraphics{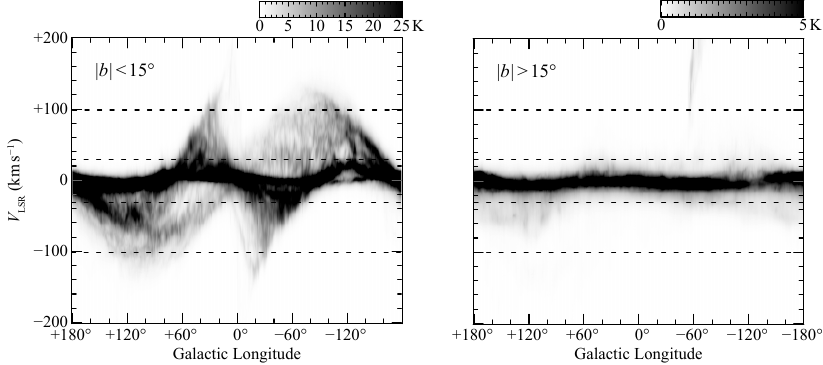}
\caption{
Galactic longitude-LSR velocity diagrams of the \ion{H}{i} in $|b|<15\degr$ (left) and $|b|>15\degr$ (right) from the HI4PI data \citep{2016A&A...594A.116H}.
The intensity is averaged over Galactic latitude.
The horizontal dashed lines in each panel indicate $V_{\mathrm{LSR}}=+100$, $+30$, $-30$, and $-100$\,km\,s$^{-1}$ (from upper to lower).
} \label{fig:pvdiagrams}
\end{figure*}

\section{Datasets}
\label{sec:datasets}

We used archival \ion{H}{i} data from the HI4PI full-sky survey \citep{2016A&A...594A.116H} and \textit{Planck} PR2 dust data \citep[]{planck2016-XLVIII}.
Both are on the \textsc{HEALPix}\footnote{\url{http://healpix.sourceforge.net/}} \citep{2005ApJ...622..759G} projection, and we performed the following analyses without reprojection.
The \textsc{HEALPix} is a framework for the pixelization of the data on the sphere.
Unlike general regular grid maps, the resolution of a \textsc{HEALPix} map is defined by the resolution parameter $N_{\mathrm{side}}$ (generally a power of 2).
The map has $12{N_{\mathrm{side}}}^{2}$ pixels of the same area over the whole sky, i.e., each pixel has a solid angle of $4\pi/(12{N_{\mathrm{side}}}^2)$\,sr.
See section 4 of \citet{2005ApJ...622..759G} for further details, and refer to table 1 in the same publication for the correspondence between $N_{\mathrm{side}}$ and mean pixel spacing.

\subsection{\ion{H}{i} data}
\label{subsec:HI_data}

The HI4PI data combines those from the first release of the Effelsberg-Bonn \ion{H}{i} Survey \citep[EBHIS,][]{2016A&A...585A..41W} and the third revision of the Galactic All-Sky Survey \citep[GASS,][]{2015A&A...578A..78K}.
The brightness temperature noise level is $\sim 43$\,mK (rms) at a velocity resolution of 1.49\,km\,s$^{-1}$.
The velocity coverage (with respect to the Local Standard of Rest, LSR) is $|V_\mathrm{LSR}|\leq 600$\,km\,s$^{-1}$ (EBHIS part) or $|V_\mathrm{LSR}|\leq 470$\,km\,s$^{-1}$ (GASS part).
The FWHM angular resolution of the combined map is 16.2\,arcmin, twice better than the Leiden/Argentine/Bonn (LAB) Survey \citep{2005A&A...440..775K}.
The data are divided into 19 parts, and each of them is presented as a FITS binary table containing spectra on the \textsc{HEALPix} grid with $N_\mathrm{side}=1024$ (the mean pixel spacing is 3.4\,arcmin).

\subsection{Dust optical depth data}
\label{subsec:tau353_data}

\citet{planck2016-XLVIII} used \textit{Planck} 2015 data release (PR2) maps and separated Galactic thermal dust emission from cosmic infrared background anisotropies by implementing the generalized needlet internal linear combination (GNILC) method.
The GNILC dust maps have a variable angular resolution with an effective beam FWHM varying from 5 to 21.8\,arcmin \citep[see fig.~2 of][]{planck2016-XLVIII}. 
The authors then produced the dust optical depth, temperature, and spectral index maps by fitting a modified blackbody model to the GNILC dust maps at 353, 545, 857\,GHz, and \textit{IRAS} 100\,{\micron} map.
We used the $\tau_{353}$ data released version R2.01 in the \textsc{HEALPix} format with $N_\mathrm{side}=2048$ (the mean pixel spacing is 1.7\,arcmin).
The median relative uncertainty in $\tau_{353}$ is $\sigma(\tau_{353})/\tau_{353}=0.037$ in $|b|>15\degr$.

Note that \citetalias{2017ApJ...838..132O}, \citetalias{2019ApJ...878..131H}\footnote{The description in section~2.3 of \citetalias{2019ApJ...878..131H}, `released version R1.10 are used', is to be corrected to `released version R1.20 are used' (Hayashi, private communication). }, \citetalias{2017PASJ...69L...5F}, \citetalias{2019ApJ...871...44T} and \citetalias{2021PASJ...73S.117F} used \textit{Planck} 2013 data release (PR1) dust data released version R1.20 \citep[]{planck2013-p06b}, whereas \citet{2019ApJ...884..130H} and the present study used the \citet{planck2016-XLVIII} data.
We compared the two datasets and found a good consistency between them (see Appendix~\ref{sec:PR2_vs_PR1}).

\subsection{Pre-processing of the data}
\label{subsec:preprocessing}
Figure~\ref{fig:pvdiagrams} shows the Galactic longitude-LSR velocity diagrams for \ion{H}{i} in both low latitudes ($|b|<15\degr$) and middle/high latitudes ($|b|>15\degr$).
The diagrams reveal the Galactic-rotational motion in the former while indicating negligibly observable rotation in the latter.
Therefore, we used the LSR velocity to define the IVCs in the middle/high latitudes rather than a deviation velocity\footnote{The deviation velocity is the difference between the velocity of a cloud and the nearest limit of the velocity allowed by a model of the galactic rotation toward the direction, introduced by \citet{1991A&A...250..499W} for low latitudes.}.
Figure~\ref{fig:pvdiagrams} also demonstrates that the LV components in the middle/high latitude are concentrated around $V_{\mathrm{LSR}}=0$\,km\,s$^{-1}$ and fall well within a range of $|V_{\mathrm{LSR}}|<30$\,km\,s$^{-1}$.
\citetalias{2015ApJ...798....6F} reported that more than 80 per cent of \ion{H}{i} in $|b|>15\degr$ are in a $V_{\mathrm{LSR}}$ range from $-9$ to $+6$\,km\,s$^{-1}$ and have $1\sigma$ velocity dispersion of smaller than 10\,km\,s$^{-1}$ \citepalias[see section 2.2 of][]{2015ApJ...798....6F}.
We set the boundary between the LV and the intermediate-velocity components at $|V_{\mathrm{LSR}}|=30$km\,s$^{-1}$.
There is no explicit definition of the boundary between the IVCs and the HVCs, and values such as $|V_{\mathrm{LSR}}|=90$\,km\,s$^{-1}$ \citep[e.g.,][]{2001ApJS..136..463W,2022MNRAS.513.3228L} or 100\,km\,s$^{-1}$ \citep[e.g.,][]{2016A&A...596A..94R,2021ApJ...923...50F} are usually adopted.
We have opted for a boundary value of 100\,km\,s$^{-1}$ in the present study.
A straight cut in the LSR velocity frame can misclassify the objects having velocities near the boundary in some directions of the sky.
It causes a loss of accuracy in the dust-to-\ion{H}{i} ratio measurements, but we have not considered the loss in the present study.

The \ion{H}{i} data were integrated into the five velocity ranges (summarized in Table~\ref{tab:velocity_components}) and combined into a single \textsc{HEALPix} format data for each velocity range.
The HI4PI \ion{H}{i} and the \textit{Planck} dust data have different resolutions, and we matched them as follows: (1) if the resolution of the \textit{Planck} data is 21.8\,arcmin at a data point, the \ion{H}{i} data were smoothed, (2) otherwise (5.0, 8.7, 11.7, or 15.0\,arcmin), the dust data were smoothed to 16.2\,arcmin resolution.
Then both were degraded to $N_\mathrm{side}=256$ (the mean pixel spacing is 13.7\,arcmin) to reduce the computational cost.
In the following sections, we used these processed data unless otherwise noted.

\section{Neutral gas and dust optical depth properties}
\label{sec:properties}

\begin{figure*}
\includegraphics{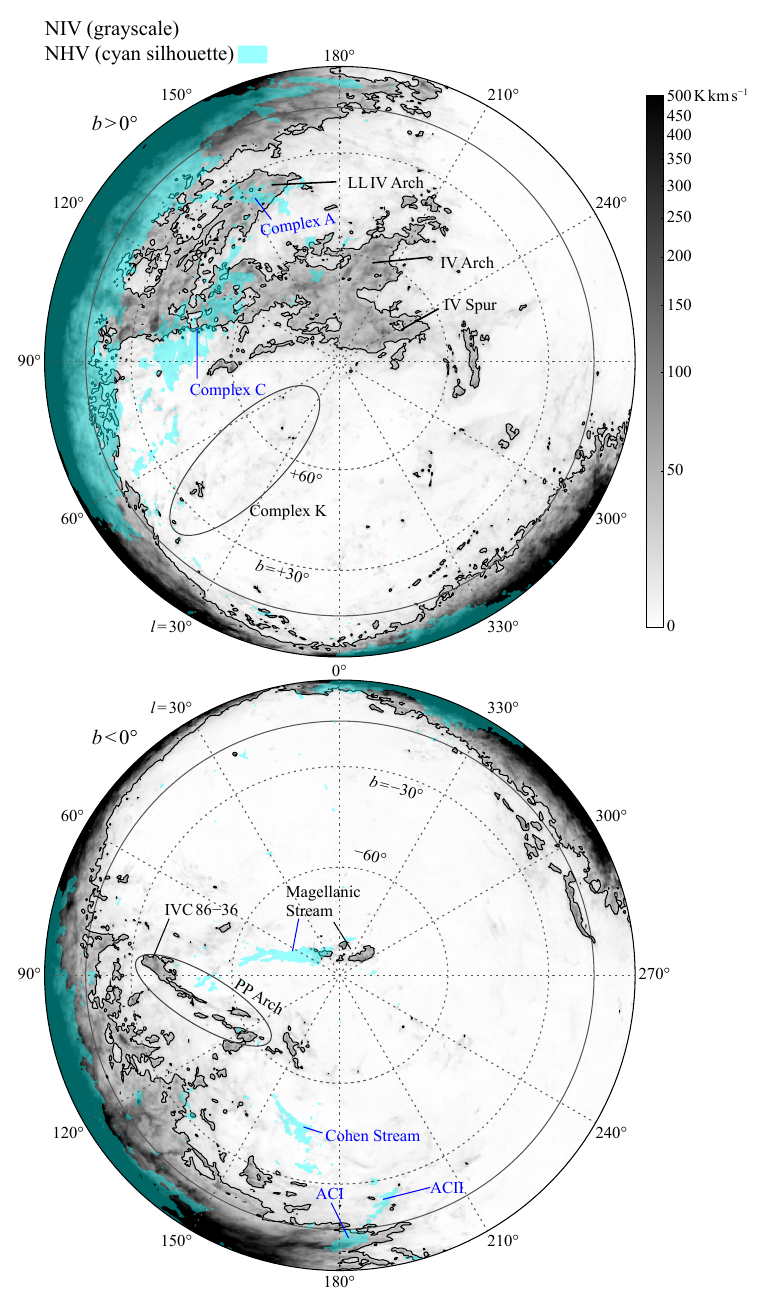}
\caption{
(a) The greyscale image shows the spatial distribution of \ion{H}{i} integrated intensity in the NIV velocity range ($V_\mathrm{LSR}= -100$--$-30$\,km\,s$^{-1}$, see Table~\ref{tab:velocity_components}), $W_{\ion{H}{i}, \mathrm{NIV}}$, in a zenithal equal area (ZEA) projection \citep{2016A&A...594A.116H}.
The contours show $W_{\ion{H}{i}, \mathrm{NIV}}=30$\,K\,km\,s$^{-1}$.
The overlaid cyan silhouette shows the NHV components ($V_\mathrm{LSR}= -470$--$-100$\,km\,s$^{-1}$) with $W_{\ion{H}{i}, \mathrm{NHV}} >10$\,K\,km\,s$^{-1}$.
Upper panels cover $b > 0\degr$ and longitude increases clockwise.
Lower panels cover $b < 0\degr$ and longitude increases counterclockwise.
Both join at $(l, b)=(0\degr, 0\degr)$.
The grey solid lines indicate $|b|=15\degr$.
} \label{fig:HI_maps}
\end{figure*}

\begin{figure*}
\begin{center}
\includegraphics{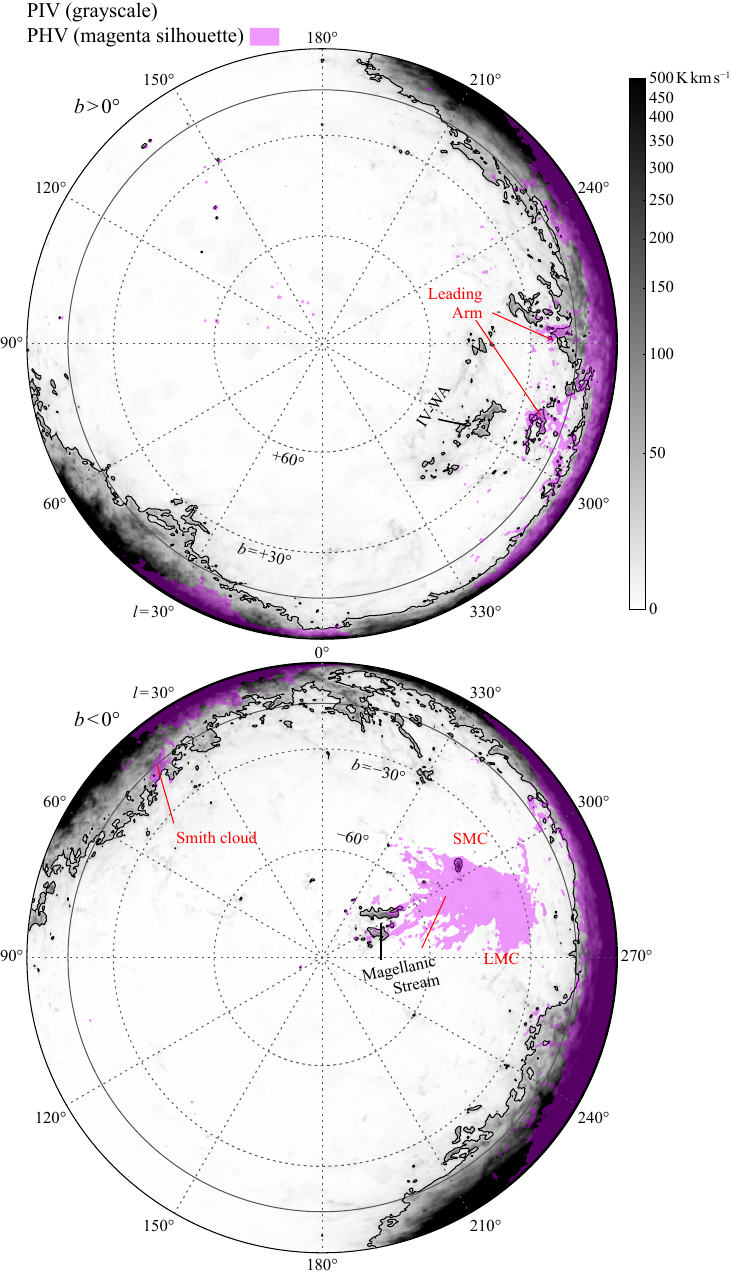}
\end{center}
\contcaption{
(b) The greyscale image shows the PIV components ($V_\mathrm{LSR}= +30$--$+100$\,km\,s$^{-1}$) and the contours show $W_{\ion{H}{i}, \mathrm{PIV}}=30$\,K\,km\,s$^{-1}$.
The overlaid magenta silhouette shows the PHV components ($V_\mathrm{LSR}= +100$--$+470$\,km\,s$^{-1}$) with $W_{\ion{H}{i}, \mathrm{PHV}} >10$\,K\,km\,s$^{-1}$.
}
\end{figure*}

\begin{figure*}
\begin{center}
\includegraphics{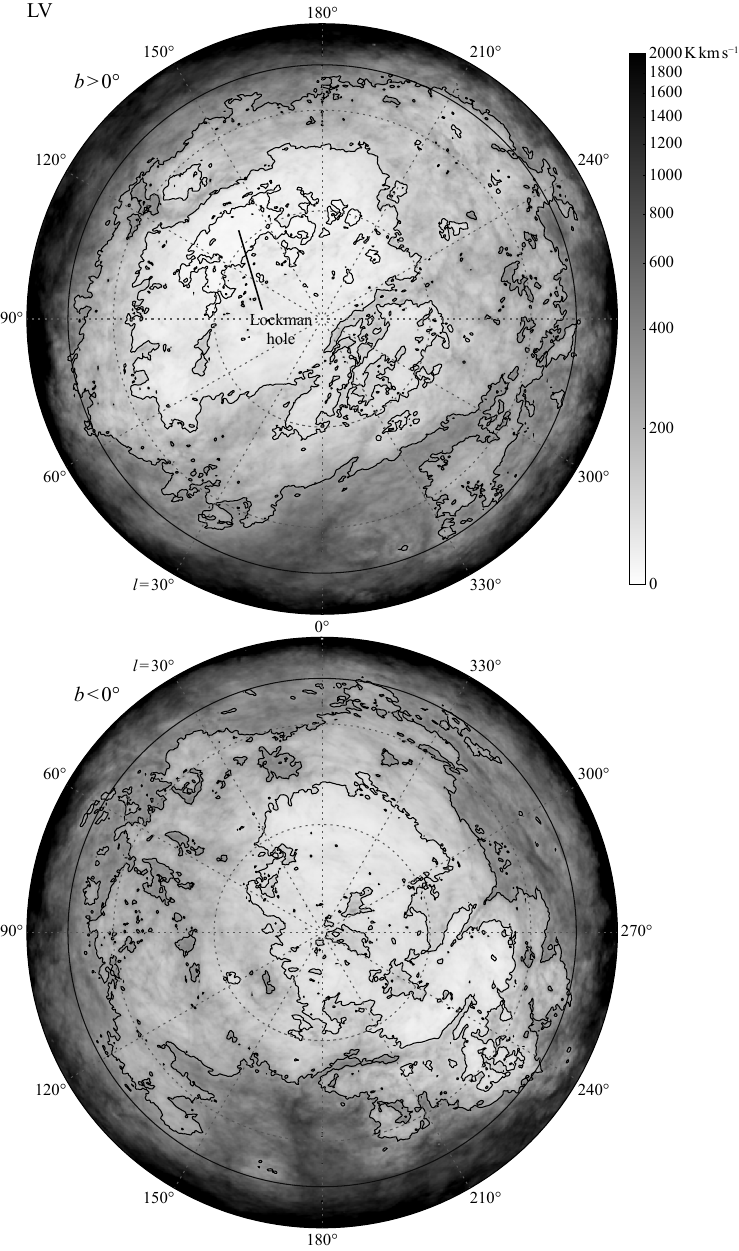}
\end{center}
\contcaption{
(c) The greyscale image shows the LV components ($V_\mathrm{LSR}= -30$--$+30$\,km\,s$^{-1}$) and the contours show $W_{\ion{H}{i}, \mathrm{LV}}=30$, 100, and 300\,K\,km\,s$^{-1}$.
}
\end{figure*}

\begin{figure*}
\includegraphics{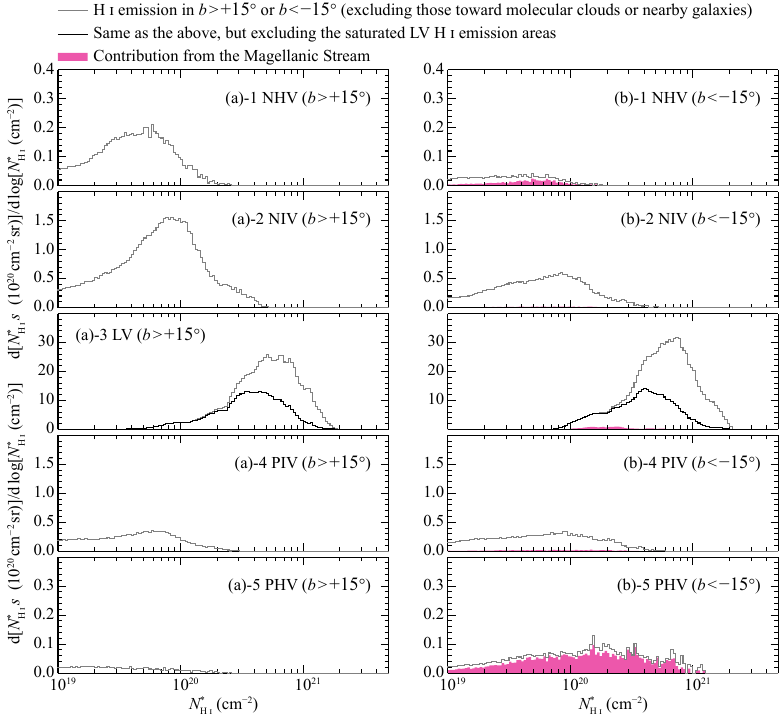}
\caption{
Distribution of the apparent amount of \ion{H}{i} gas (the product of the column density and the solid angle, $N^{\ast}_{\ion{H}{i}, X}s$) as a function of the column density under the optically thin approximation $N_{\ion{H}{i}}^{\ast}$ (equation~(\ref{eqn:columndensity})) in the NHV, NIV, LV, PIV and PHV velocity ranges (from top to bottom), excluding toward molecular clouds or nearby galaxies (see Section~\ref{subsec:masking}).
The panels in the left column are for $b>+15\degr$, and those in the right column are for $b<-15\degr$.
The black lines in panels (a)-3 and (b)-3 further exclude the saturated LV \ion{H}{i} emission (see Section~\ref{subsec:regressionanalysis}).
The shades in magenta show the contribution from the Magellanic Stream.
} \label{fig:nhi_histo}
\end{figure*}

\begin{table*}
	\caption{Summary of the velocity ranges in the present study.}
	\label{tab:velocity_components}
	\begin{threeparttable}
	\begin{tabular}{l@{ }lr@{ -- }rrrr}
		\hline
		 & & \multicolumn{2}{c}{$V_\mathrm{LSR}$ range\tnote{b}} & $N_{\ion{H}{i}}^{\ast}s$ ($|b|>15\degr$)\tnote{c} & \multicolumn{2}{c}{$N_{\ion{H}{i}}^{\ast}s$ (with confident $\widehat{\zeta}$)\tnote{d}} \\
		\cline{6-7}
		\multicolumn{2}{c}{Name\tnote{a} } & \multicolumn{2}{c}{(km\,s$^{-1}$)} & \multicolumn {1}{c}{(cm$^{-2}$\,sr)} & \multicolumn {1}{c}{(cm$^{-2}$\,sr)} \\
	\hline
		Negative high velocity & (NHV) & $-470$ & $-100$ & $2\times 10^{19}$ & $2\times 10^{18}$ & 10\% \\
		Negative intermediate velocity & (NIV) & $-100$ & $-30$ & $2\times 10^{20}$ & $5\times 10^{19}$ & 28\% \\
		Low velocity & (LV) & $-30$ & $+30$ & $4\times 10^{21}$ & $2\times 10^{21}$ & 40\% \\
		Positive intermediate velocity & (PIV) & $+30$ & $+100$ & $7\times 10^{19}$ & $9\times 10^{18}$ & 13\% \\
		Positive high velocity & (PHV) & $+100$ & $+470$ & $5\times 10^{18}$ & $3\times 10^{17}$ & 6\% \\
	\hline
	\end{tabular}
	\begin{tablenotes}
	\item[a] Identification name and its abbreviation form in parentheses.
	\item[b] Minimum and maximum velocities with respect to the LSR.
	\item[c] The total apparent amount of \ion{H}{i} gas obtained by equation~(\ref{eqn:HI_amount}) in $|b|>15\degr$ skys excluding the Magellanic Stream.
	\item[d] The total apparent amount of \ion{H}{i} gas of the pixels with $\sigma(\widehat{\zeta})/\zeta_{\mathrm{soln}}$ better than 0.1 and the percentage of above.
	\end{tablenotes}
	\end{threeparttable}
\end{table*}

Fig.~\ref{fig:HI_maps} shows the spatial distributions of $W_{\ion{H}{i}}$ in the negative (i.e., blue-shifted) intermediate velocity (NIV, $V_\mathrm{LSR}=-100$--$-30$\,km\,s$^{-1}$), positive (red-shifted) intermediate velocity (PIV, $V_\mathrm{LSR}=+30$--$+100$\,km\,s$^{-1}$), and LV ($|V_\mathrm{LSR}|<30$\,km\,s$^{-1}$) components, respectively.
Those of the negative- and positive high-velocity (NHV and PHV, $|V_\mathrm{LSR}|>100$\,km\,s$^{-1}$) components are also shown as silhouettes.
Fig.~\ref{fig:nhi_histo} shows the distribution function of the apparent amount (the product of the column density and the solid angle) of \ion{H}{i} gas,
\begin{equation}\label{eqn:HI_amount}
\sum_{i} N_{\ion{H}{i}, X}^{\ast}(l_{i}, b_{i})s
\end{equation}
as a function of the column density,
\begin{equation}\label{eqn:columndensity}
N_{\ion{H}{i}, X}^{\ast}(l_{i}, b_{i}) = C_{0}\,W_{\ion{H}{i}, X}(l_{i}, b_{i}),
\end{equation}
where $X=\mathrm{NHV}$, NIV, $\cdots$, PHV represents the velocity ranges listed in Table~\ref{tab:velocity_components}, ($l_{i}$, $b_{i}$) are the galactic coordinates of the $i$-th pixel ($i=1, 2, \cdots$), and $C_{0}=1.82\times 10^{18}\,(\mathrm{cm}^{-2}\,\mathrm{K}^{-1}\,\mathrm{km}^{-1}\,\mathrm{s})$.
The solid angle of each pixel is given by $s=4\pi / (12{N_{\mathrm{side}}}^{2})$\,sr, and $N_\mathrm{side}=256$ in the present dataset.
The asterisk put on $N_{\ion{H}{i}}$ means the \ion{H}{i} column density is under the optically-thin approximation (optical depth $\tau_{\ion{H}{i}}\ll 1$), following the notation of \citetalias{2014ApJ...796...59F}, \citetalias{2015ApJ...798....6F}, \citetalias{2017ApJ...838..132O} and \citetalias{2019ApJ...878..131H}.
As reported in the previous works, the IVCs are mainly in the negative velocity range.
The NIV components are concentrated in giant IVC complexes Intermediate-Velocity (IV)~Arch, IV~Spur, Low-Latitude~(LL)~IV~Arch, and Complex~M occupying one-third of $b>15\degr$ skies.
Another prominent IVC, PP~Arch, is located in the Galactic South and has a head-tail structure elongated from IVC~86~$-$36 to $(l, b)=(150\degr, -60\degr)$.
The only small apparent amount of \ion{H}{i} gas is in the positive velocity ranges, except for the Magellanic System.
The LV components covering the whole sky are considered to be the local volume gas within 300\,pc of the sun.
They have an \ion{H}{i} column density range of $10^{20}$--$10^{21}$\,cm$^{-2}$ with a peak at 3--$4\times 10^{20}$\,cm$^{-2}$, which is one order of magnitude higher than IVCs ($1\times 10^{19}$ to $2\times 10^{20}$\,cm$^{-2}$) and HVCs ($\lesssim 1\times 10^{20}$\,cm$^{-2}$).

\begin{figure*}
\includegraphics{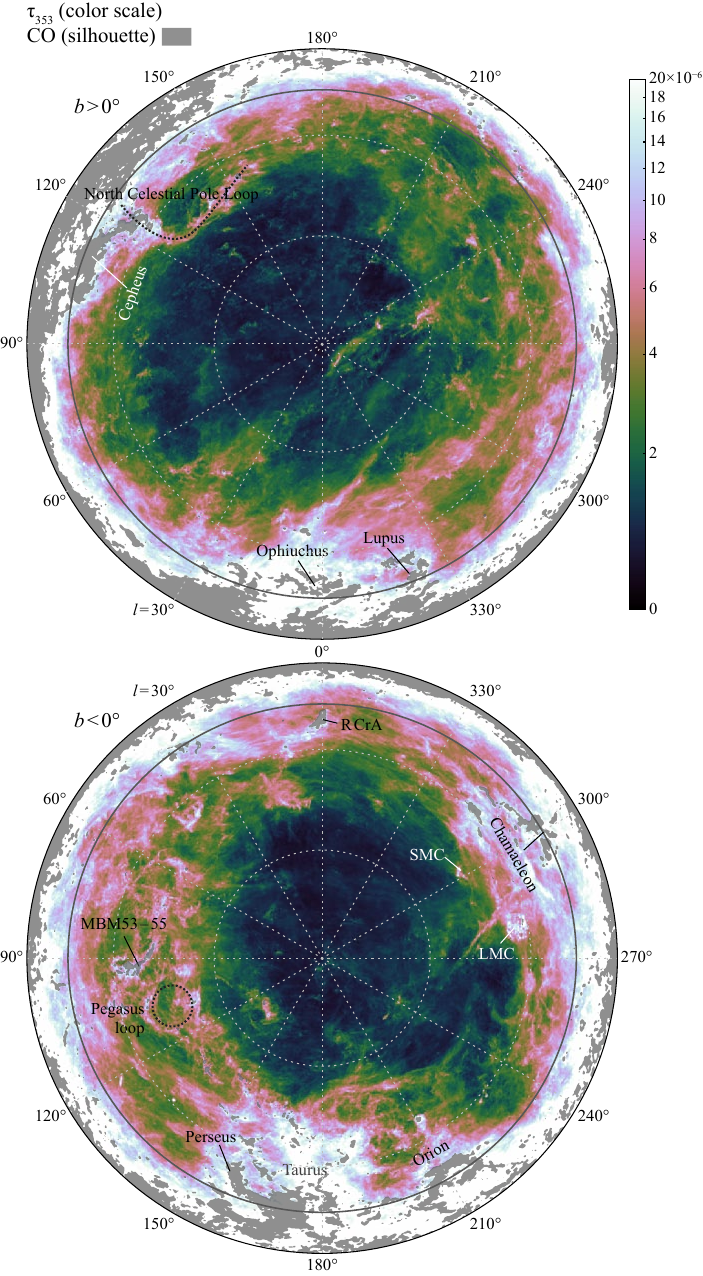}
\caption{
Spatial distribution of 353\,GHz dust optical depth ($\tau_{353}$) \citep{planck2016-XLVIII} shown in the same projection as Fig.~\ref{fig:HI_maps}. 
The overlaid silhouette outlines the molecular clouds with $W_\mathrm{CO}>1.4$\,K\,km\,s$^{-1}$ \citep{planck2014-a12}.
} \label{fig:tau353_maps}
\end{figure*}

\begin{figure*}
\begin{center}
\includegraphics{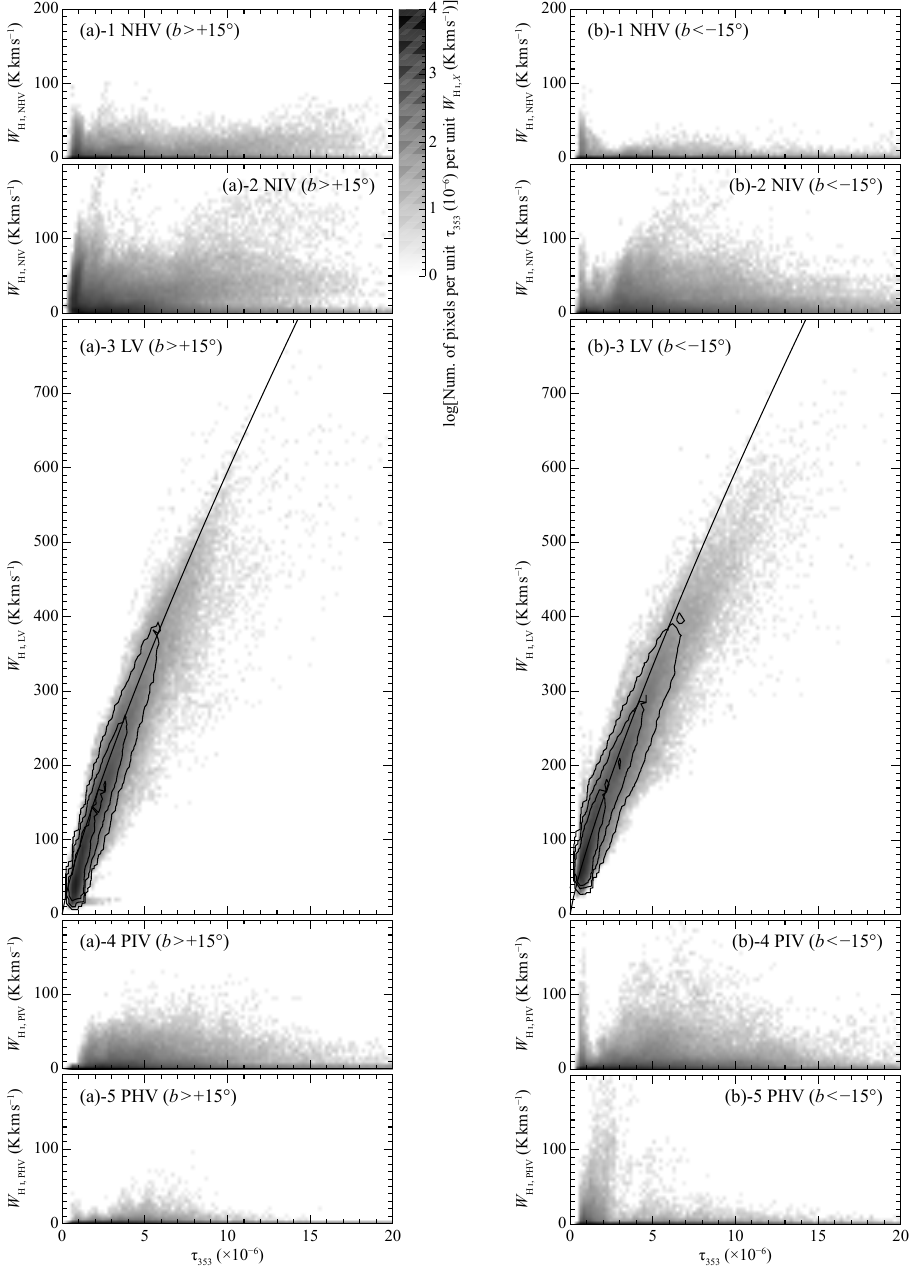}
\end{center}
\caption{
Density plots showing the correlation between $\tau_{353}$ and $W_{\ion{H}{i}}$ in the NHV, NIV, LV, PIV and PHV velocity ranges (from top to bottom), excluding toward molecular clouds or nearby galaxies (see Section~\ref{subsec:masking}).
The plots in panels (a)-3 and (b)-3 further exclude the saturated LV \ion{H}{i} (see Section~\ref{subsec:regressionanalysis}).
The $W_{\ion{H}{i}}$ values are obtained by integrating within each velocity range, but $\tau_{353}$ values are the total amount on the line-of-sights (not velocity-decomposed).
The panels in the left column are for $b>+15\degr$ and those in the right column are for $b<-15\degr$.
The contours panels (a)-3 and (b)-3 contain 50, 75 and 90 per cent of data points.
The solid lines indicate the solar-neighbourhood dust-to-\ion{H}{i} ratio ($\zeta_{\mathrm{soln}}$) line in this work $W_{\ion{H}{i}}\,(\mbox{K\,km\,s$^{-1}$})=(1.59\times 10^{25})/(1.82\times 10^{18})\times {\tau_{353}}^{1/1.2}$ (see Section~\ref{sec:analyses}).
}\label{fig:global_correlation}
\end{figure*}

Fig.~\ref{fig:tau353_maps} shows the spatial distribution of $\tau_{353}$, and Fig.~\ref{fig:global_correlation} shows the correlation plots between $\tau_{353}$ and $W_{\ion{H}{i}}$ in the $|b| > 15\degr$ skies with masking described in Section~\ref{subsec:masking}.
Here, the $W_{\ion{H}{i}}$ values are velocity-decomposed, but the $\tau_{353}$ values are not.
The good correlation between the two quantities is found only in the LV component, while the other components show a rather poor correlation.
The plots demonstrate that the LV component dominates $\tau_{353}$ in almost every single direction, consistent with the remarkably similar spatial distribution of the LV \ion{H}{i} gas (Fig.~\ref{fig:HI_maps}(c)) and $\tau_{353}$ (Fig.~\ref{fig:tau353_maps}).
Fig.~\ref{fig:global_correlation} also presents nearly vertical features displaced or broadened along the $\tau_{353}$ axis in the high- and intermediate-velocity components.
It indicates that dust-poor HV/IV \ion{H}{i} overlaps with the dust-rich LV \ion{H}{i}.

\section{The Estimate of the dust-to-\ion{H}{i} ratio}
\label{sec:analyses}

\begin{figure}
\begin{center}
\includegraphics{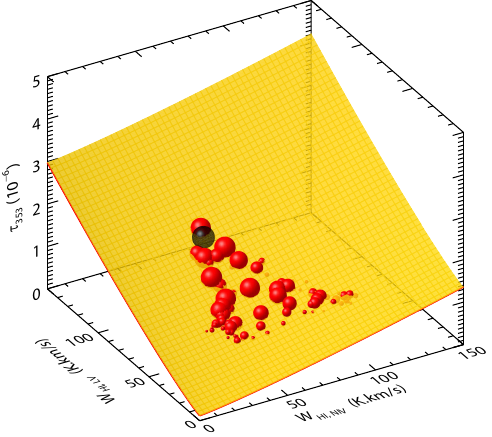}
\end{center}
\caption{
A toy model of the regression analysis.
The 3D scatter plot in the $W_{\ion{H}{i}, \mathrm{NIV}}$-$W_{\ion{H}{i}, \mathrm{LV}}$-$\tau_{353}$ space shows the data points at and in the neighbour of a regression point $(l, b)=(150\fdg 1, +67\fdg 6)$ (the black sphere and the red ones, respectively).
The radius of each sphere is proportional to the 1/3 power of the weight given by equation (\ref{eqn:weighting_function}) (decays with the angular distance in the sky from the regression point, maximum is 1.0 and truncated below $10^{-3}$).
The $W_{\ion{H}{i}, \mathrm{NHV}}$, $W_{\ion{H}{i}, \mathrm{PIV}}$, and $W_{\ion{H}{i}, \mathrm{PHV}}$ are small enough for these data points, and the contribution from these components to $\tau_{353}$ can be regarded as zero.
The best-fit surface given by equation (\ref{eqn:regression_model_libi}) with $\widehat{\zeta}_{\mathrm{NIV}}(150\fdg 1, +67\fdg 6)/\zeta_{\mathrm{soln}}=0.70\pm 0.09$ and $\widehat{\zeta}_{\mathrm{LV}}(150\fdg 1, +67\fdg 6)/\zeta_{\mathrm{soln}}=1.40\pm 0.08$ is also shown.
} \label{fig:3d_plots}
\end{figure}

The present work uses $W_{\ion{H}{i}}$ and $\tau_{353}$ to derive the dust-to-\ion{H}{i} ratio in the LV, IVC and HVC components separately.
The basic concept introduced in the previous studies \citepalias{2017PASJ...69L...5F,2019ApJ...871...44T,2021PASJ...73S.117F} is that the regression coefficient of a single regression between $W_{\ion{H}{i}}$ and $\tau_{353}$ (or the gradient of the best-fit line in the $W_{\ion{H}{i}}$-$\tau_{353}$ plane) gives the dust-to-\ion{H}{i} ratio.
However, in the framework of the present study, $\tau_{353}$ is the sum of the contribution from the five velocity components, and it is impossible to estimate each ratio by a single regression between, e.g., $W_{\ion{H}{i}, \mathrm{NIV}}$ and a velocity-decomposed  $\tau_{353, \ion{H}{i}, \mathrm{NIV}}$.
Assuming that $\tau_{353}$ is a linear combination of multiple $W_{\ion{H}{i}, X}$-terms (see Section~\ref{subsec:formulation}), we can estimate the dust-to-\ion{H}{i} ratios of all components simultaneously as the partial coefficients of a multiple regression, or the gradient of the best-fit ``plane'' in an ($m+1$)-dimensional space, where $m$ is the number of the $W_{\ion{H}{i}, X}$ terms and $m=5$ in the present study.

The dust-to-\ion{H}{i} ratio of each velocity component is not uniform across the sky, and we, therefore, used the geographically weighted regression (GWR) technique \citep{10.1111/j.1538-4632.1996.tb00936.x,Fotheringham2002} which allows us to derive the spatial distribution of the dust-to-\ion{H}{i} ratio, whereas the general regression estimates a set of global and spatially-invariant coefficients.
The GWR technique is an outgrowth of the ordinary least-squares (OLS) regression.
It is an advanced extension of the moving-window technique and estimates local coefficients at each pixel using a distance-decay weighting function.

Figure \ref{fig:3d_plots} is a visualization of the regression-analysis process at a regression point (a pixel where we want to estimate the local coefficients) ($l$, $b$)=($150\fdg 1$, $+67\fdg 6$), in a three-dimensional ($W_{\ion{H}{i}, \mathrm{LV}}$-$W_{\ion{H}{i}, \mathrm{NIV}}$-$\tau_{353}$) space\footnote{We made the regressions between the five \ion{H}{i} terms and $\tau_{353}$, as described in the following subsections, but $W_{\ion{H}{i}, \mathrm{NHV}}$, $W_{\ion{H}{i}, \mathrm{PIV}}$ and $W_{\ion{H}{i}, \mathrm{PHV}}$ are negligible in many cases.}.
We find that the ``plane''\footnote{The best-fit ``plane'' is slightly curved due to the small non-linearity of the sub-mm emission by the dust growth as expressed by an exponent $\alpha=1.2$ (see Section~\ref{subsec:formulation}).} immediately gives the two gradients $\tau_{353}$-$W_{\ion{H}{i}, \mathrm{LV}}$ and $\tau_{353}$-$W_{\ion{H}{i}, \mathrm{NIV}}$, and these gradients give the dust-to-\ion{H}{i} ratios of the LV and NIV component, respectively.

As presented in Section~\ref{sec:properties}, $\tau_{353}$ is dominated by the LV component, with $\sim 1$--2 orders smaller contributions from the other components.
One might wonder whether extracting such faint and reduced contributions is possible.
The issue could be severe if we subtract the contribution from the LV component and analyze the remaining, as done in \citetalias{2021PASJ...73S.117F}.
In the present method, if we explain using the above model, the gradients in the $W_{\ion{H}{i}, \textrm{NIV}}$ and $W_{\ion{H}{i}, \textrm{LV}}$ directions are mutually independent, and it does not matter that the LV is dominant in $\tau_{353}$.
The uncertainty of $\tau_{353}$ ($\sigma(\tau_{353})$, typically on the order of $10^{-8}$ in the directions analyzed) probably provides the extraction limit in the present method.
The HVCs, Magellanic Stream and some IVCs (such as PP-Arch) are close to the limit (typically estimated to be $\mathrm{several}\times 10^{-8}$), whereas the IVC complexes IV~Arch/Spur have large enough $\tau_{353}$ ($\gtrsim 10^{-7}$) (see also Section~\ref{subsec:regressionanalysis}).

\subsection{Masking}
\label{subsec:masking}
\begin{figure}
\begin{center}
\includegraphics{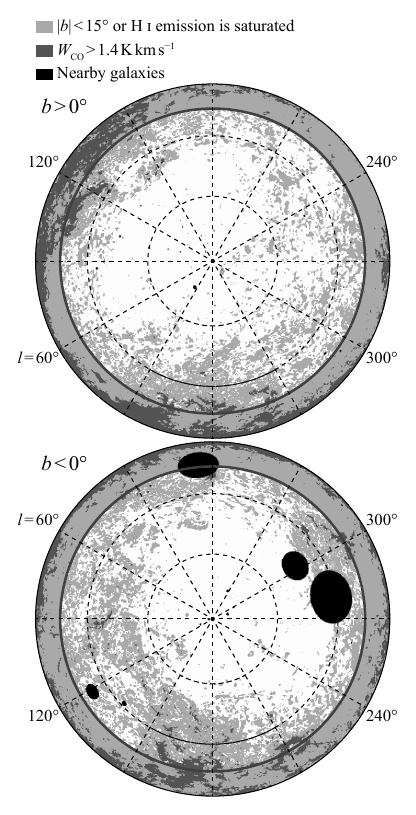}
\end{center}
\caption{
The masked and saturated \ion{H}{i} areas in the present study (Sections~\ref{subsec:masking} and \ref{subsec:regressionanalysis}) shown in the same projection as Fig.~\ref{fig:HI_maps}.
The low latitude zone $|b|<15\degr$ is delimited by thick solid lines.
The dark-grey shadow outlines the area with $W_\mathrm{CO}>1.4$\,K\,km\,s$^{-1}$ \citep{planck2014-a12}.
The black-filled ellipses cover nearby galaxies taken from the catalogue by \citet{2013AJ....145..101K} and samples studied by \citet{2016MNRAS.460.2143W}.
The light-grey shadow in $|b|>15\degr$ presents the areas where LV \ion{H}{i} emission is saturated.
} \label{fig:maskmaps}
\end{figure}

\begin{figure}
\begin{center}
\includegraphics{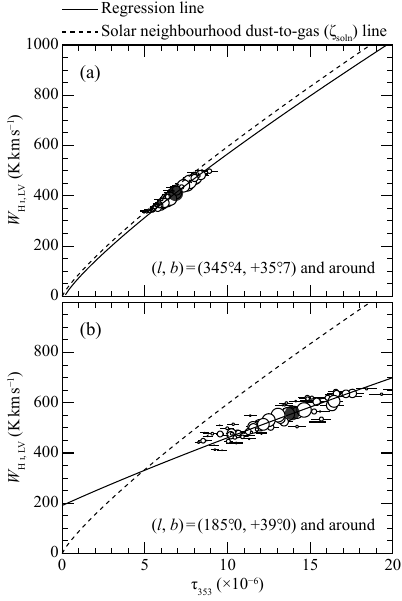}
\end{center}
\caption{
(a) Scatter plot showing the $\tau_{353}$-$W_{\ion{H}{i}, \mathrm{LV}}$ correlation for data points at and in the neighbour (within 1.2\,deg) of a regression point $(l, b)=(345\fdg 4, +35\fdg 7)$ where the LV \ion{H}{i} emission is judged to be optically thin (not  saturated).
The filled circle is the data point at the regression point, and the open circles are those in the neighbour.
The radius of each circle is proportional to the 1/2 power of the weight given by equation (\ref{eqn:weighting_function}), (decays with the angular distance in the sky from the regression point, maximum is 1.0 and truncated below $10^{-3}$).
The horizontal bars attached to the data points show $\sigma(\tau_{353})$ \citep{planck2016-XLVIII}.
The solid lines show local regression lines described by equation~(\ref{eqn:regression_model_libi}) at the position.
The dashed line indicates the solar-neighbourhood dust-to-\ion{H}{i} ratio ($\zeta_{\mathrm{soln}}$) line in this work (see Section~\ref{sec:analyses}).
(b) Same as (a) but for data points within 1.2\,deg of $(l, b)=(185\fdg 0, -39\fdg 0)$ where the LV \ion{H}{i} emission is judged to be saturated.
In both samples, the \ion{H}{i} integrated intensity of the NHV, NIV, PIV, and PHV components are small enough ($W_{\ion{H}{i}, X}<5.5$\,K\,km\,s$^{-1}$ or $N_{\ion{H}{i}, X}^{\ast}<1\times 10^{19}$\,cm$^{-2}$), and their contribution to $\tau_{353}$ is approximately zero.
} \label{fig:partial_correlation}
\end{figure}

The pixels which meet any of the criteria (a)--(c) were masked before the regression analysis;
\begin{enumerate}
\renewcommand{\labelenumi}{(\alph{enumi})}
\item Galactic latitude is $|b|<15\degr$ in order to eliminate contamination by the Galactic-disc components far from us.
\item
CO emission is detected at the $3\sigma$ level (1.4\,K\,km\,s$^{-1}$), where the molecular gas is not negligible compared to the atomic gas.
We used the \textit{Planck} PR2 type 2 CO(1--0) map \citep{planck2014-a12}.
The median uncertainty of the map is $\sigma(W_\mathrm{CO})=0.44$\,K\,km\,s$^{-1}$ in $|b|>15\degr$. 
\item The areas covering nearby galaxies listed in either or both the catalogue by \citet{2013AJ....145..101K} and the samples studied by \citet{2016MNRAS.460.2143W}.
\end{enumerate}
Fig.~\ref{fig:maskmaps} summarises the masked areas.

\subsection{Formulation}
\label{subsec:formulation}

The observed $\tau_{353}$ is the sum of the contribution from each component,
\begin{equation} \label{eqn:tau353_sum_of_contributions}
\begin{split}
\tau_{353}(l_{i}, b_{i}) = & \sum_{X} \tau_{353, \mathrm{H}^{+}, X}(l_{i}, b_{i}) + \sum_{X} \tau_{353, \ion{H}{i}, X}(l_{i}, b_{i}) \\ 
 & +\sum_{X} \tau_{353, \mathrm{H}_{2}, X}(l_{i}, b_{i}),
\end{split}
\end{equation}
where $\tau_{353, \mathrm{H}^{+}, X}$, $\tau_{353, \ion{H}{i}, X}$, and $\tau_{353, \mathrm{H}_{2}, X}$ are the contribution from the dust associated with ionized, atomic, and molecular gas in the velocity range $X$, respectively. 
The molecular fraction can be approximated to be zero in the unmasked regions.
If we assume that the contribution from the ionized component is negligibly small (see also discussion in Section~\ref{subsec:ionizedgas}), then equation~(\ref{eqn:tau353_sum_of_contributions}) can be rewritten as
\begin{equation} \label{eqn:tau353_sum_of_contributions_approx}
\tau_{353}(l_{i}, b_{i}) = \sum_{X} \tau_{353, \ion{H}{i}, X}(l_{i}, b_{i}).
\end{equation}
The contribution from a component $X$ is expressed as a function of its \ion{H}{i} column-density $N_{\ion{H}{i}, X}$, introducing a dust-to-\ion{H}{i} ratio parameter $\zeta$,
\begin{equation} \label{eqn:totalcolumn_to_tau353}
\tau_{353, \ion{H}{i}, X}(l_{i}, b_{i}) = \left[ \zeta_{X}(l_{i}, b_{i}) N_{\ion{H}{i}, X}(l_{i}, b_{i}) \right]^{\alpha}.
\end{equation}
We used a $N_{\ion{H}{i}}$ model having a nonlinear relationship with $\tau_{353}$ found by \citetalias{2017ApJ...838..132O} and \citetalias{2019ApJ...878..131H}.
These authors used the 21\,cm \ion{H}{i} data with $\tau_{353}$ following \citetalias{2014ApJ...796...59F} and \citetalias{2015ApJ...798....6F} by taking into account the optical depth effect of the 21\,cm \ion{H}{i} emission.
\citetalias{2017ApJ...838..132O} derived a $\tau_{353}$-$N_{\ion{H}{i}}$ relationship with $\alpha = 1.3$ for the \ion{H}{i} gas in the Perseus region, and \citetalias{2019ApJ...878..131H} obtained $\alpha = 1.2$ in the Chamaeleon molecular cloud complex.
The value of $\alpha$ greater than 1.0 was suggested to be due to the dust evolution effect by \citet{2013ApJ...763...55R} who derived the non-linearity with $\alpha=1.3$ from (the far infrared optical depth)-(near infrared colour excess) relationship in Orion.
\citet{2019ApJ...884..130H} made a \textit{Fermi}-LAT $\gamma$-ray analysis and confirmed the non-linearity with $\alpha\sim 1.4$.
In the following analyses, we adopted $\alpha=1.2$.

The \ion{H}{i} column density is a function of $W_{\ion{H}{i}, X}$ and \ion{H}{i} optical depth $\tau_{\ion{H}{i}, X}$, 
\begin{equation} \label{eqn:HI_intensity_to_column}
N_{\ion{H}{i}, X}(l_{i}, b_{i}) = C_{0}\frac{\tau_{\ion{H}{i}, X}(l_{i}, b_{i})}{1-\exp\left[ -\tau_{\ion{H}{i}, X}(l_{i}, b_{i}) \right]}W_{\ion{H}{i}, X}(l_{i}, b_{i}).
\end{equation}
As it is not feasible to simultaneously estimate both $\zeta_{X}(l_{i}, b_{i})$ and $\tau_{\ion{H}{i}, X}(l_{i}, b_{i})$ as free parameters in a regression model due to the compounded relationship arising from their multiplication, we assume that
\begin{equation} \label{eqn:HI_intensity_to_column_approx}
N_{\ion{H}{i}, X}(l_{i}, b_{i}) \sim N_{\ion{H}{i}, X}^{\ast}(l_{i}, b_{i}) = C_{0}W_{\ion{H}{i}, X}(l_{i}, b_{i})
\end{equation}
under the optically thin approximation of the \ion{H}{i} emission ($\tau_{\ion{H}{i}, X} \ll 1$)  (see Section~\ref{subsec:regressionanalysis} for exception handling for optically thick cases).
Then, we set up the regression equation using equations~(\ref{eqn:tau353_sum_of_contributions_approx}), (\ref{eqn:totalcolumn_to_tau353}) and (\ref{eqn:HI_intensity_to_column_approx}) as 
\begin{equation} \label{eqn:regression_model_libi}
\tau_{353}(l_{i}, b_{i}) = \tau_{353, 0}(l_{i}, b_{i}) + {C_{0}}^{\alpha} \sum_{X} \left[ \zeta_{X}(l_{i}, b_{i}) W_{\ion{H}{i}, X}(l_{i}, b_{i}) \right]^{\alpha}.
\end{equation}
The constant term $\tau_{353, 0}(l_{i}, b_{i})$ is the amount of $\tau_{353}$ unaffected by variation in $W_{\ion{H}{i}, X}$, e.g., the zero-level offsets (see also discussion in Section~\ref{subsec:ionizedgas}), and has been found empirically to take far off from zero if \ion{H}{i} emission is saturated due to the optical depth effect (Fig.~\ref{fig:partial_correlation}).

\subsection{Regression analysis and resulsts}
\label{subsec:regressionanalysis}

\begin{figure*}
\includegraphics{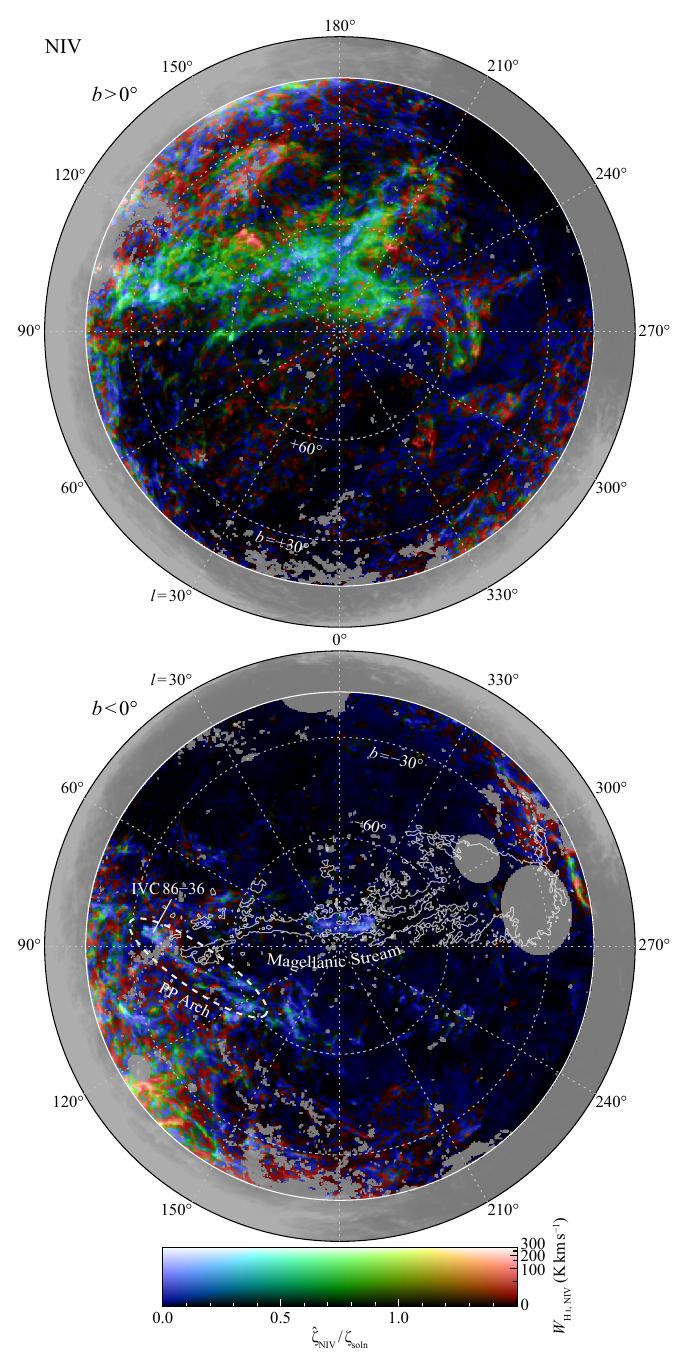}
\caption{
(a) Spatial distribution of the relative dust-to-\ion{H}{i} ratio $\widehat{\zeta}_{\mathrm{NIV}}/\zeta_{\mathrm{soln}}$ encoded as colour hue and $W_{\ion{H}{i}, \mathrm{NIV}}$ presented by brightness, shown in the same projection as Fig.~\ref{fig:HI_maps}.
The grey shadow presents the areas with no valid $\widehat{\zeta}_{\mathrm{NIV}}$ values.
The white contours in the lower panel outline the Magellanic Stream.
} \label{fig:distribution_of_zeta}
\end{figure*}

\begin{figure*}
\begin{center}
\includegraphics{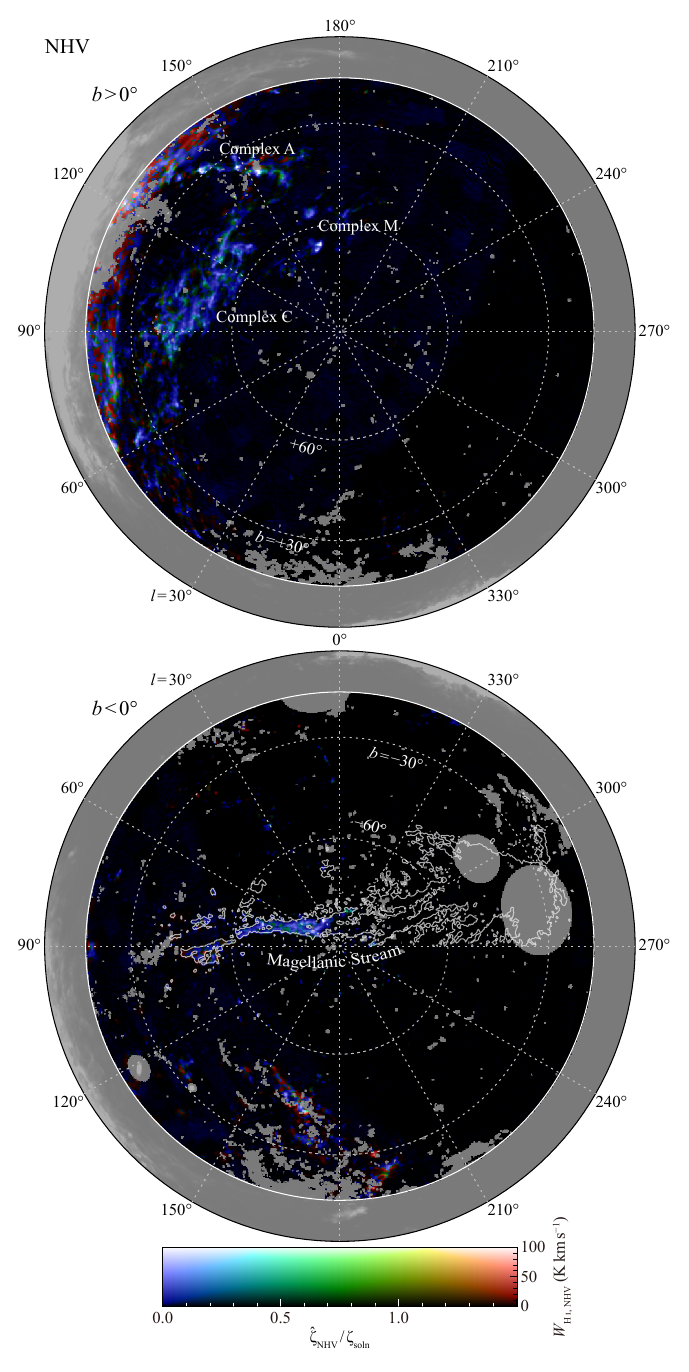}
\end{center}
\contcaption{
(b) Same as (a) but for the NHV.
}
\end{figure*}

\begin{figure*}
\begin{center}
\includegraphics{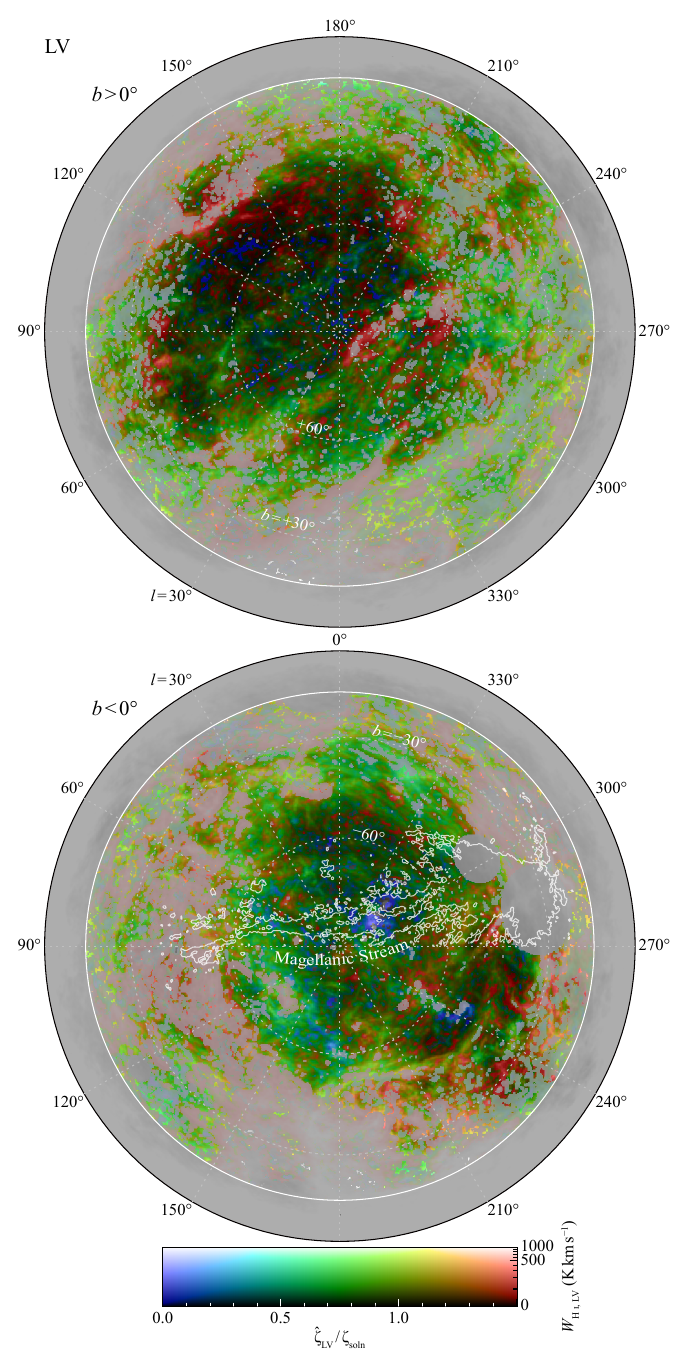}
\end{center}
\contcaption{
(c) Same as (a) but for the LV.
}
\end{figure*}

\begin{figure*}
\begin{center}
\includegraphics{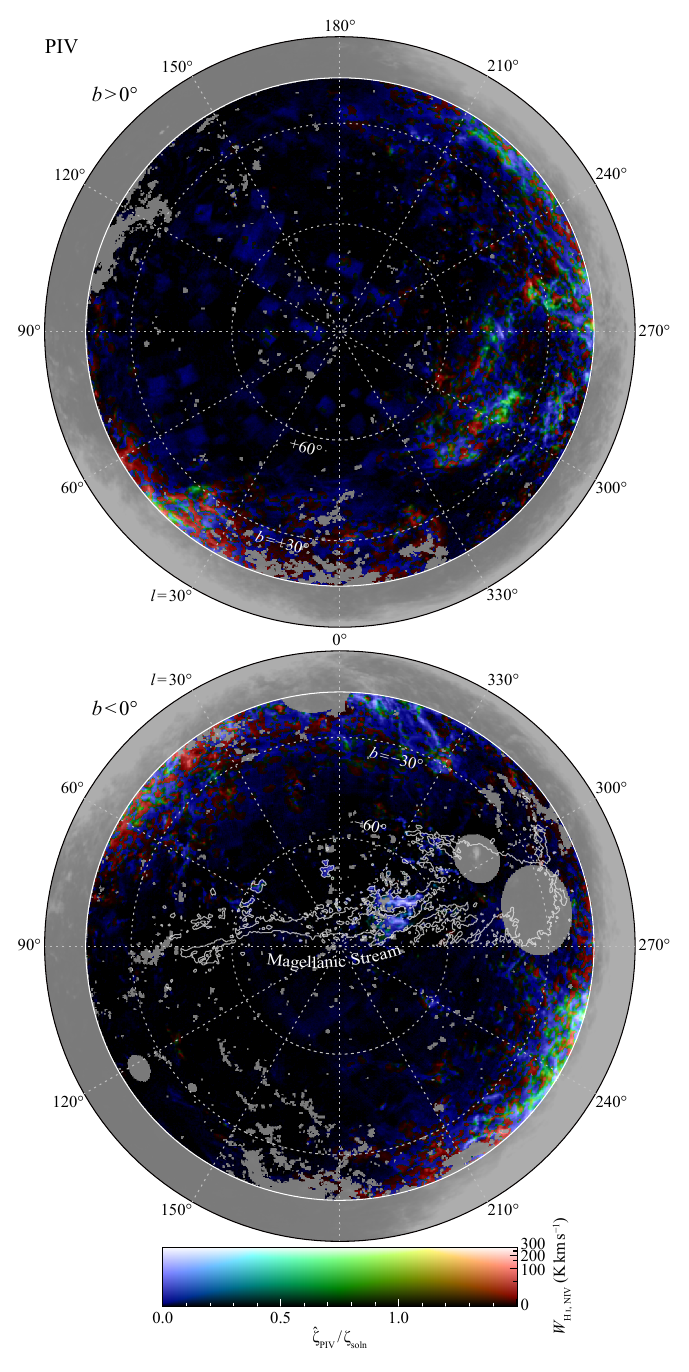}
\end{center}
\contcaption{
(d) Same as (a) but for the PIV.
}
\end{figure*}

\begin{figure*}
\begin{center}
\includegraphics{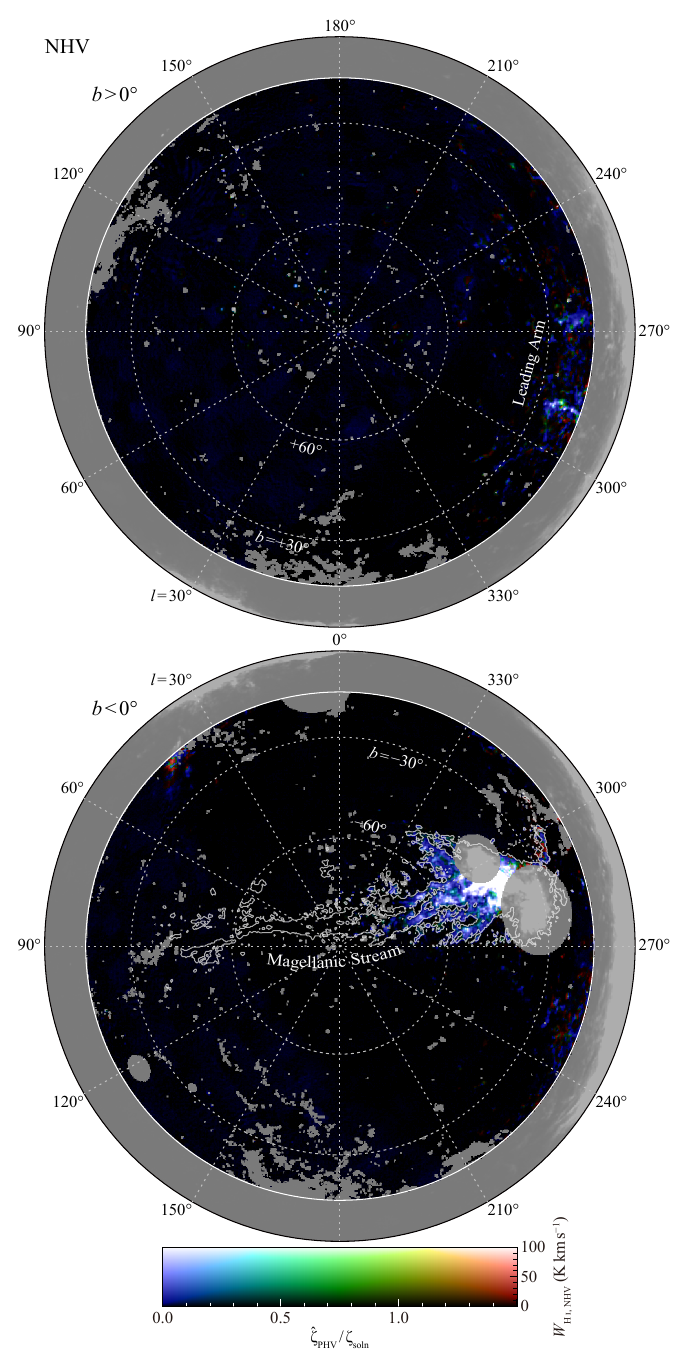}
\end{center}
\contcaption{
(e) Same as (a) but for the PHV.
}
\end{figure*}

\begin{figure*}
\begin{center}
\includegraphics{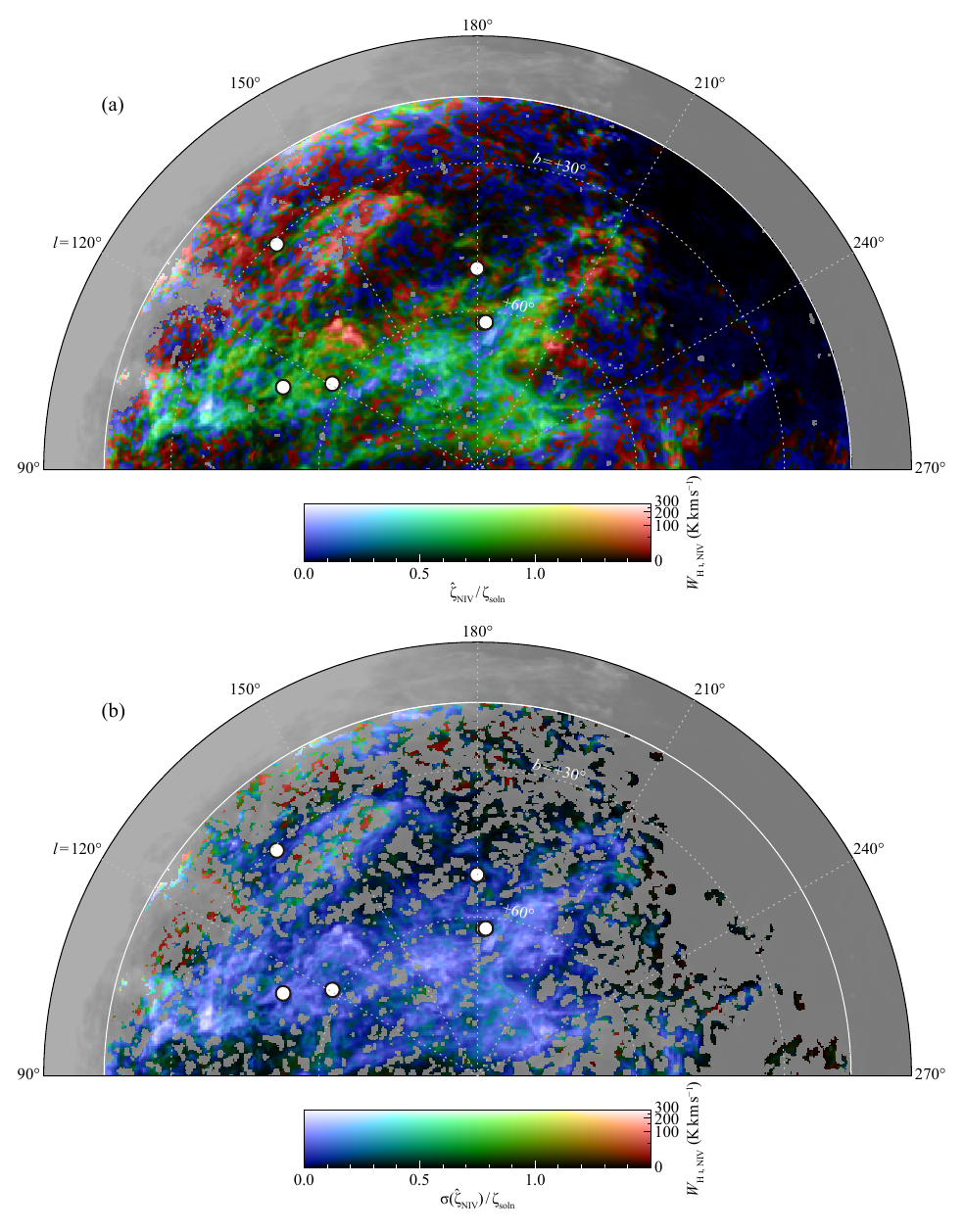}
\end{center}
\caption{
(a) A close up view of Fig.~\ref{fig:distribution_of_zeta}(a) focusing on the NIV-GN region containing IV~Arch/Spur, LL~IV~Arch, and Complex~M.
(b) Same as (a) but the colour hue shows the standard error $\sigma(\widehat{\zeta}_\mathrm{NIV})/\zeta_{\mathrm{soln}}$.
The filled circles indicate the positions of the background objects where absorption measurements were made (see Table \ref{tab:list_abs_measure}).
The grey shadow presents the areas with no valid $\widehat{\zeta}_{\mathrm{NIV}}$ or $\sigma(\widehat{\zeta}_{\mathrm{NIV}})$ values.
} \label{fig:zeta+sezeta_IVArch}
\end{figure*}

\begin{figure*}
\begin{center}
\includegraphics{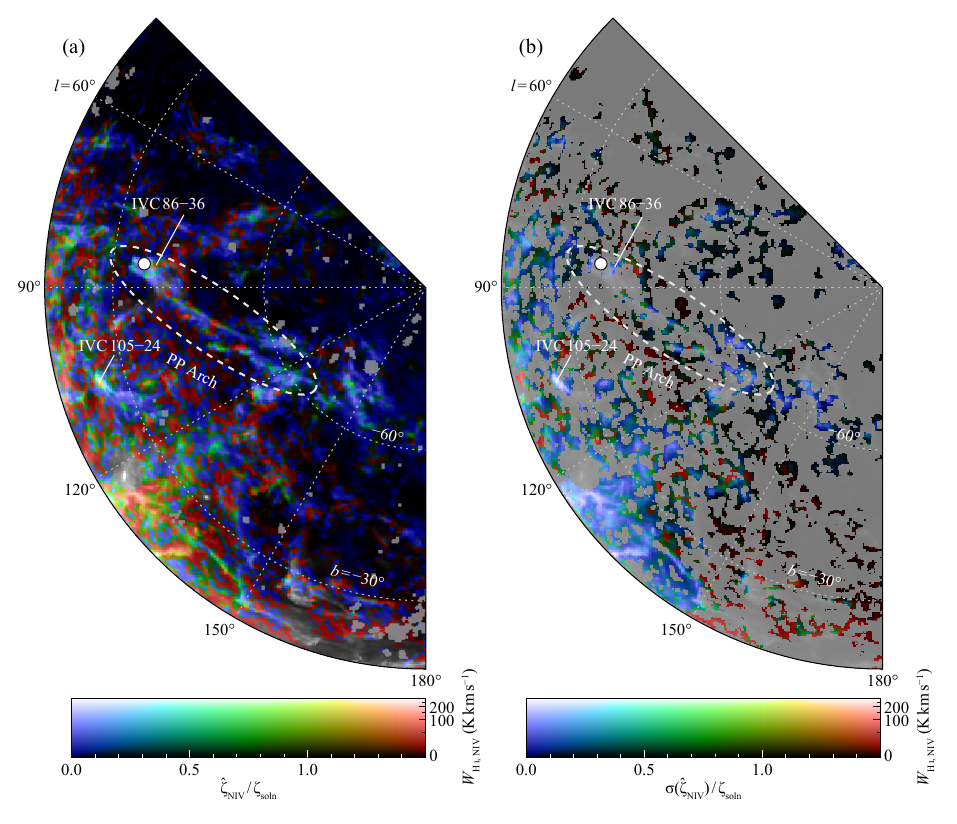}
\end{center}
\caption{
Same as Figure \ref{fig:zeta+sezeta_IVArch} but focusing on the NIV-GS region including PP~Arch.
} \label{fig:zeta+sezeta_PPArch}
\end{figure*}

\begin{figure*}
\begin{center}
\includegraphics{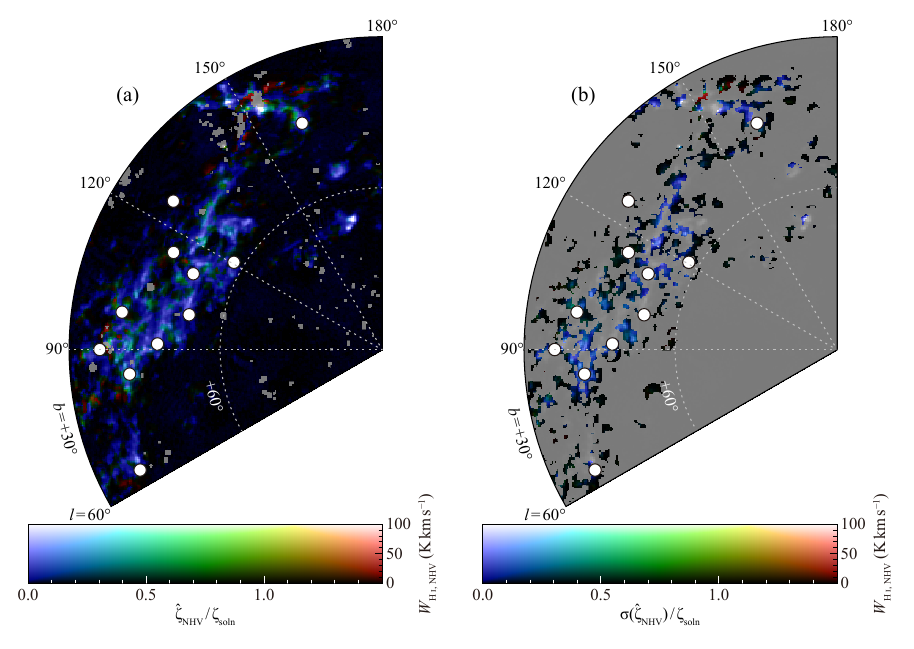}
\end{center}
\caption{
(a) A close up view of Fig.~\ref{fig:distribution_of_zeta}(b) focusing on the NHV-GN region including Complexes~A/C/M.
(b) Same as (a) but the colour hue shows the standard error $\sigma(\widehat{\zeta}_\mathrm{NHV})/\zeta_{\mathrm{soln}}$.
The filled circles indicate the positions of the background objects where absorption measurements were made (see Table \ref{tab:list_abs_measure}).
The grey shadow presents the areas with no valid $\widehat{\zeta}_{\mathrm{NHV}}$ or $\sigma(\widehat{\zeta}_\mathrm{NHV})$ values.
} \label{fig:zeta+sezeta_ComplexC}
\end{figure*}

\begin{figure*}
\begin{center}
\includegraphics{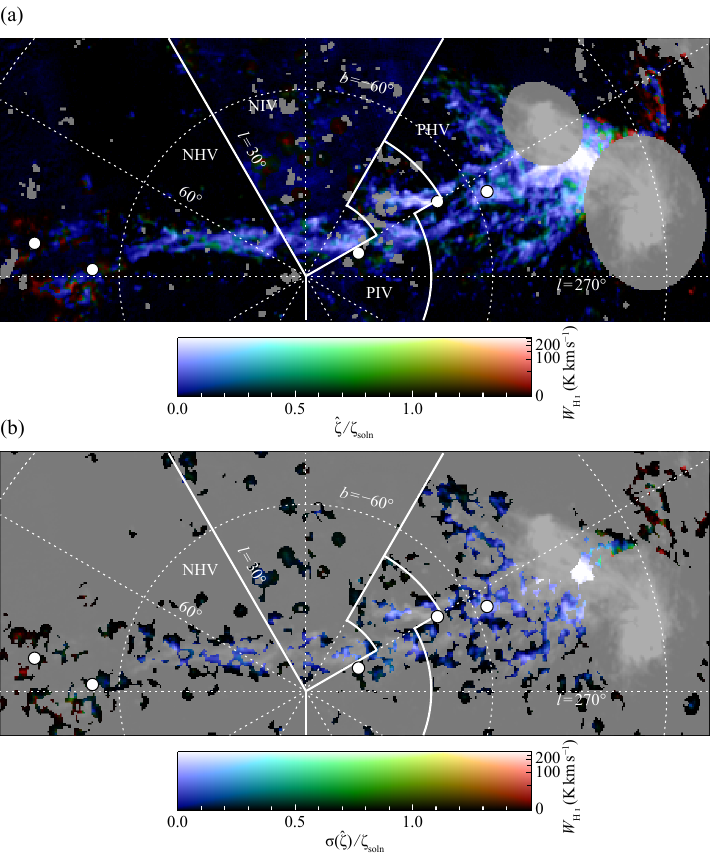}
\end{center}
\caption{
(a) A patchwork of the NHV, NIV, PIV, and PHV maps showing the spatial distribution of the dust-to-\ion{H}{i} ratio $\widehat{\zeta}/\zeta_{\mathrm{soln}}$ (colour) and $W_{\ion{H}{i}, X}$ (brightness) in the Magellanic Stream.
(b) Same as (a) but the colour hue shows the standard error $\sigma(\widehat{\zeta})/\zeta_{\mathrm{soln}}$.
The filled circles indicate the positions of the background objects where absorption measurements were made (see Table \ref{tab:list_abs_measure}).
The grey shadow presents the areas with no valid $\widehat{\zeta}$ or $\sigma(\widehat{\zeta})$ values.
} \label{fig:zeta+sezeta_MS}
\end{figure*}

\begin{figure}
\begin{center}
\includegraphics{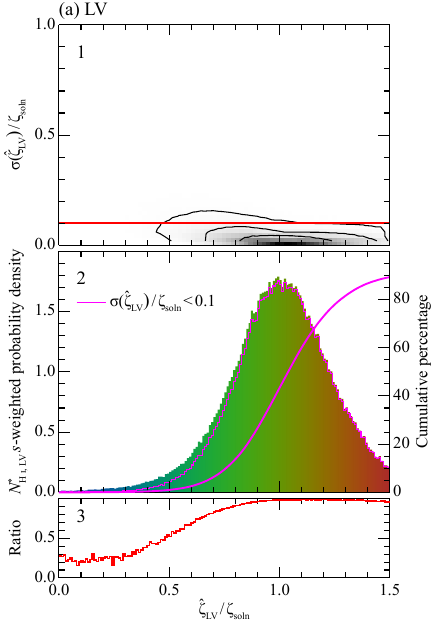}
\end{center}
\caption{
(a)-1 Density plot showing the two-dimensional probability density with respect to $\widehat{\zeta}_\mathrm{LV}/\zeta_{\mathrm{soln}}$ (horizontal axis) and its standard error $\sigma(\widehat{\zeta}_\mathrm{LV})/\zeta_{\mathrm{soln}}$ (vertical axis) for the LV component, weighted by the apparent amount of \ion{H}{i} gas $N^{\ast}_{\ion{H}{i}, \mathrm{LV}}s$ of each pixel.
The contours contain 50, 75, and 90 per cent of the total amount.
The horizontal line indicates $\sigma(\widehat{\zeta}_\mathrm{LV})/\zeta_{\mathrm{soln}}=0.1$.
(a)-2 The filled histogram shows (i) a projection of the above onto the horizontal axis, and the solid line histogram shows (ii) the subset better than $\sigma(\widehat{\zeta}_\mathrm{LV})/\zeta_{\mathrm{soln}}=0.1$.
The overlaid curves depict the cumulative percentage of (ii) (the scale is displayed on the right-hand vertical axis).
(a)-3 The ratio of (ii) to (i).
} \label{fig:zeta_distribution}
\end{figure}

\begin{figure*}
\begin{center}
\includegraphics[scale=.97]{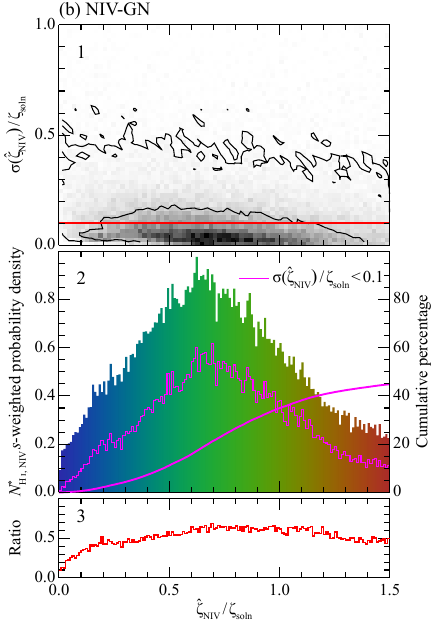}
\includegraphics[scale=.97]{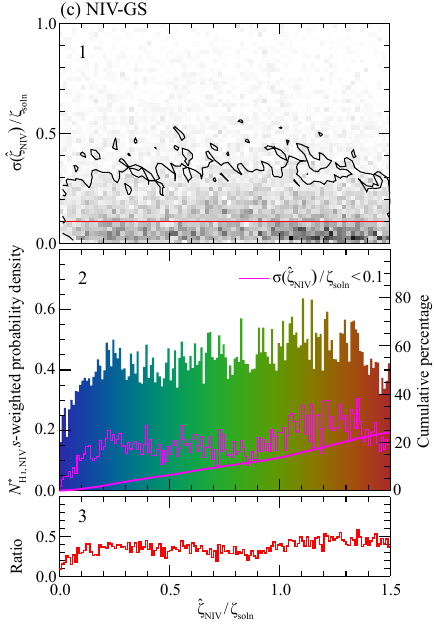}
\includegraphics[scale=.97]{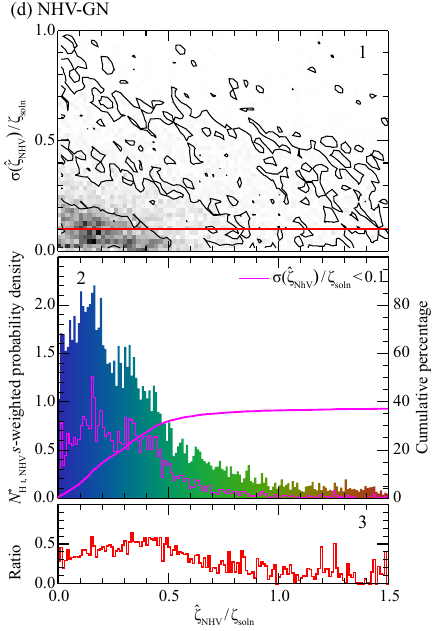}
\includegraphics[scale=.97]{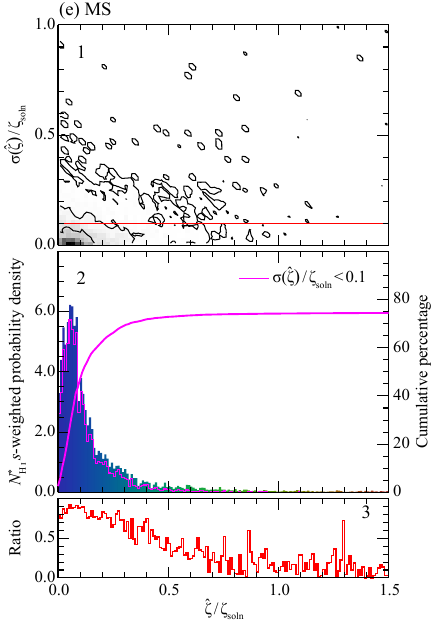}
\end{center}
\contcaption{
Same as (a) but for (b) the NIV-GN, (c) NIV-GS, (d) NHV-GN, and (e) MS regions.
}
\end{figure*}

We estimated $\zeta(l_{i}, b_{i})$ at each pixel using the GWR technique.
We used a truncated-Gaussian weighting function\footnote{It is known empirically that the results of GWR are relatively insensitive to the choice of weighting function \citep[e.g.,][]{Fotheringham2002}.
Not only Gaussian but also, for example, bi-square functions are often used.}
\begin{equation} \label{eqn:weighting_function}
w_{i}(l_{j}, b_{j})=\left\{\begin{array}{ll}
	\exp\left[-\displaystyle{\frac{1}{2}}\left(\frac{\theta_{ij}}{\theta_\mathrm{bw}}\right)^{2}\right] & (\theta_{ij} \leq \theta_\mathrm{tr}) \\
	0 & (\mbox{otherwise})
\end{array}\right.,
\end{equation}
where $\theta_{ij}$ is the great-circle angular distance between the regression point and the $j$-th ($j=1, \cdots, n$) pixel $(l_{j}, b_{j})$, $\theta_\mathrm{bw}$ is the bandwidth of the weighting function,  and $\theta_\mathrm{tr}$ is the truncation radius.
We used $\theta_\mathrm{bw}=20$\,arcmin ($\mathrm{FWHM}=47$\,arcmin), and $\theta_\mathrm{tr}=1.2$\,degree (truncates below $w_{i}(l_{j}, b_{j})=1\times 10^{-3}$).
The $\zeta$ cannot take negative values; negative $\zeta$ does not make physical sense and will overestimate the other coefficients.
We thus introduced a non-negative least-squares (NNLS) regression technique.
In the case either or both of the dust-to-\ion{H}{i} ratio and the column density of the component $X$ is meagre at a regression point, i.e., $\widehat{\tau}_{353, \ion{H}{i}, X} = (\widehat{\zeta}_{X} C_{0}W_{\ion{H}{i}, X})^{\alpha}$\footnote{Throughout this study, the symbols with hat ( $\widehat{\phantom{x}}$ ) are estimates from regressions.} is roughly to say on the order of $10^{-8}$ or less, $\zeta_{X}$ is force set to 0 by using the algorithm of \citet{Lawson1995} (see their section~2 of chapter~23, also briefly summarised in Appendix \ref{subsec:GWR_main} of the present paper) and treated as having no valid standard error value.
The procedures and formulae in the regression analysis are summarised in Appendix \ref{sec:GWR}.

Although the regression model in equation (\ref{eqn:regression_model_libi}) assumes optically-thin \ion{H}{i} emission, saturation due to the optical-depth effects has been observed even in $|b|>15\degr$ skys \citepalias{2015ApJ...798....6F}.
The regression coefficients in saturation cases do not accurately represent the dust-to-\ion{H}{i} ratio; therefore, we excluded them from the subsequent analysis using a simple judgment method.
Fig.~\ref{fig:partial_correlation} shows the $\tau_{353}$-$W_{\ion{H}{i}, \mathrm{LV}}$ correlations for optically thin (not saturated) and saturated examples.
The best-fit line for the latter takes a large constant term (or intercept), whereas the one for the former throughs near the origin.
We determined that \ion{H}{i} emission toward a direction is saturated if the estimated value of the constant term $|\widehat{\tau}_{353, 0}|$ exceeds $1\times 10^{-6}$ ($1\sigma$ of all analysed pixels).
Fig.~\ref{fig:maskmaps} shows the saturated \ion{H}{i} area.
Note that saturation should not be an issue for the IVCs and HVCs due to their low column densities.
This procedure was applied only to the LV component.

Fig.~\ref{fig:distribution_of_zeta} shows the spatial distribution of estimated $\widehat{\zeta}/\zeta_{\mathrm{soln}}$, where the solar neighbourhood dust-to-\ion{H}{i} ratio $\zeta_{\mathrm{soln}}=6.29\times 10^{-26}$\,cm$^{2}$ is an empirical constant determined so that the LV component has a mode value of $\widehat{\zeta}_{\mathrm{LV}}/\zeta_{\mathrm{soln}}=1.0$ and $\widehat{\zeta}/\zeta_{\mathrm{soln}}$ present the ratios relative to the solar-neighbourhood value.
Figs.~\ref{fig:zeta+sezeta_IVArch}--\ref{fig:zeta+sezeta_MS} show close-up views of $\widehat{\zeta}/\zeta_{\mathrm{soln}}$ and their standard error $\sigma(\widehat{\zeta})/\zeta_{\mathrm{soln}}$ focusing on four regions-of-interest (ROIs),  NIV-Galactic North (GN), NIV-Galactic South (GS), NHV-GN, and the Magellanic Stream (MS).
Fig.~\ref{fig:zeta_distribution} shows the distribution functions of $\widehat{\zeta}/\zeta_{\mathrm{soln}}$ and $\sigma(\widehat{\zeta})/\zeta_{\mathrm{soln}}$ in the LV and the four ROIs for all the pixels with valid values and also the fraction determined with $\sigma(\widehat{\zeta})/\zeta_{\mathrm{soln}}$ better than 0.1.
We find that in the IVCs and HVCs in the ROIs, the fraction with significant determination is $\sim 50$ per cent.
This fraction is significantly larger than the previous results when the coverage in the sky is considered, particularly in terms of the solid angle.
We give a detailed description of the individuals in the following subsection.

\subsection{Description of the individuals}
\subsubsection{The LV component}\label{subusbsection:LV_components}
Fig.~\ref{fig:zeta_distribution}(a) shows the distribution function of $\widehat{\zeta}_{\mathrm{LV}}/\zeta_{\mathrm{soln}}$.
The ratio is determined in most pixels with a sufficiently high confidence level better than $\sigma(\widehat{\zeta}_{\mathrm{LV}})/ \zeta_{\mathrm{soln}}=0.1$.
The peak of the distribution is normalised to 1.0 as intended, and a Gaussian function well approximates the distribution.
The measurements of the metallicity of G dwarfs were made for the local volume within 25\,pc by many authors in the last five decades \citep[e.g.,][]{1996MNRAS.279..447R}.
These works indicate that the dispersion of metallicity is 0.2--0.3\,dex with a peak at $-0.2$\,dex.
The dispersion is somewhat larger than the present one.
Considering the much longer age of G dwarfs, several Gyr, than the dynamical timescale of \ion{H}{i} gas, $\sim $\,Myr, the difference seems reasonable, and the \ion{H}{i} gas in the local volume seems to be well mixed probably by the turbulent motion.

\subsubsection{The NIV-GN region}
The NIV-GN region ($l=90\degr$--$270\degr$, $b>15\degr$) contains a large apparent amount of IVC in the complexes IV~Arc/Spur, LL~IV~Arch and Complex~M (Complex M straddles the $-100$\,km\,s$^{-1}$ boundary, part of which is in the NHV range in the present study).

The $\widehat{\zeta}_{\mathrm{NIV}}/\zeta_{\mathrm{soln}}$ in this region is peaked at 0.6 with a broad continuous shape ranging from less than 0.1 to high values beyond 1.5 (Fig.~\ref{fig:zeta_distribution}(b)).
The absorption measurements toward bright background objects indicated the metallicity is near solar (see Table \ref{tab:list_abs_measure}), i.e., not much different from the LV components.
The present dust-to-\ion{H}{i} ratio distribution gives a significantly different trend; there exists a significant amount of gas having $\widehat{\zeta}_{\mathrm{NIV}}/\zeta_{\mathrm{soln}}<0.5$ (of the fractions with significance better than the $\sigma(\widehat{\zeta}_{\mathrm{NIV}})/\zeta_{\mathrm{soln}}=0.1$ threshold, 10 per cent display $\widehat{\zeta}_{\mathrm{NIV}}/\zeta_{\mathrm{soln}}<0.3$ and 20 per cent $\widehat{\zeta}_{\mathrm{NIV}}/\zeta_{\mathrm{soln}}<0.5$).
This difference is likely caused by the tiny fraction of the IVCs observed by the absorption measurements.

\subsubsection{The NIV-GS region}
The most prominent object in the NIV-GS region ($l=45\degr$--$180\degr$, $b>15\degr$)  is PP~Arch.
The present result reveals that the head of PP~Arch (IVC~86$-$36) has a low $\widehat{\zeta}_\mathrm{NIV}/\zeta_{\mathrm{soln}}$ distribution peaked at $\sim 0.2$--0.3, consistent with the \citepalias{2021PASJ...73S.117F} results, and the tail exhibits even lower values.
Another head-tail structure IVC elongated from $(l, b) = (105\degr, -24\degr)$ parallel to PP~Arch (referred to as IVC~105$-$24 for convenience) shows similar $\widehat{\zeta}_\mathrm{NIV}/\zeta_{\mathrm{soln}}$ values, suggesting that they are the same type of object.
Many other unidentified minor IVCs are in low latitudes, some of which appear to be connected to the Galactic disc.
The $\widehat{\zeta}_\mathrm{NIV}/\zeta_{\mathrm{soln}}$ values in the NIV-GS region exhibit a gently flat distribution ranging from 0.1 or below to 1.5 and above (Fig.~\ref{fig:zeta_distribution}(c)).

\subsubsection{The NHV-GN region}
The NHV-GN region ($l=60\degr$--$180\degr$, $b>30\degr$) focuses on Complex~C as well as Complexes~A and M.
Complex~C has the largest apparent size among HVC complexes in $|b|>15\degr$ skies outside the Magellanic Stream.

There have been several attempts to detect thermal dust emission associated with HVCs \citep[e.g.,][]{1986A&A...170...84W,2014MNRAS.441.2266S}, some possible detections \citep[e.g.,][]{2009ApJ...692..827P,2016A&A...586A.121L}, but no confident one have been reported\footnote{\citealt{2005ApJ...631L..57M} claims the first detection, but this is controversial, as argued by \citet[][]{2009ApJ...692..827P}.}.
The difficulty is due to the low column density and the low metallicity.
In the present work, barely sufficient column density in the scattered small areas allows the dust-to-\ion{H}{i} estimation (Fig.~\ref{fig:zeta+sezeta_ComplexC}). 
Fig.~\ref{fig:zeta_distribution}(d) indicates that $\widehat{\zeta}_{\mathrm{NHV}}/\zeta_{\mathrm{soln}}$ in this region peaked at 0.1--0.2 with a relative uncertainty of $\sigma(\widehat{\zeta}_{\mathrm{NHV}})/\widehat{\zeta}_{\mathrm{NHV}}\sim 1$ and is even lower than $\widehat{\zeta}_\mathrm{NIV}/\zeta_{\mathrm{soln}}$ in the NIV-GN region (Fig.~\ref{fig:zeta_distribution}(b)).
More than 50 per cent of the fractions with $\sigma(\widehat{\zeta}_{\mathrm{NHV}})/\zeta_{\mathrm{soln}}<0.1$ exhibit $\widehat{\zeta}_{\mathrm{NHV}}<0.3$, indicating the low dust content in the HVC.

\subsubsection{The Magellanic Stream}
The Magellanic Stream is a well-known giant filamentous structure, starting at the Magellanic Clouds and trailing over 100 degrees, having a large velocity gradient from $+180$\,km\,s$^{-1}$ near the Clouds to $-450$\,km\,s$^{-1}$ at the tip.
Fig.~\ref{fig:zeta+sezeta_MS} is a patchwork of the NHV, NIV, PIV, and PHV maps showing the spatial distribution of $\widehat{\zeta}/\zeta_{\mathrm{soln}}$ and Fig.~\ref{fig:zeta_distribution}(e) shows the distribution function.
The MS has the lowest dust-to-\ion{H}{i} ratio among the four ROIs, with a mode at $\sim 0.1$ and a tail extending to $\sim 0.5$.

\section{Discussion}
\label{sec:discussion}

\subsection{Comparison of the present dust-to-\ion{H}{i} ratio with the absorption-line measured metallicity}\label{subsec:present_vs_absorption}

\begin{table*}
	\caption{List of the absorption measurements.}
	\label{tab:list_abs_measure}
	\begin{threeparttable}
	\begin{tabular}{llrrlrrlr}
		\hline
		 & \multicolumn{7}{c}{Absorption Measurements} \\
		\cline{2-8}
		 & \multicolumn{1}{c}{Background\tnote{a}} & \multicolumn{2}{c}{Position\tnote{b}} & \multicolumn{1}{c}{Ion} & \multicolumn{1}{c}{$V_{\mathrm{LSR}}$\tnote{c}} & \multicolumn{1}{c}{$A/A_{\sun}$\tnote{d}} & \multicolumn{1}{c}{Ref.\tnote{e}} & \multicolumn{1}{c}{$\widehat{\zeta}/\zeta_{\mathrm{soln}}$\tnote{f}} \\
		 & & \multicolumn{1}{c}{$l$} & \multicolumn{1}{c}{$b$} & & \multicolumn{1}{c}{ (km\,s$^{-1}$)} \\
		\hline
		Complex~A & I~Zw~18 & $160\fdg 53$ & $+44\fdg 84$ & \ion{O}{i} & $-160$ & 0.062 & 1, 2 & $0$ \\
		\hline
		Complex~C & Mrk~501 & $63\fdg 60$ & $+38\fdg 86$ & \ion{O}{i} & $-135$--$-65$ & $<0.29$ & 3 & $0$ \\
		 & PG~1626$+$554 & $84\fdg 51$ & $+42\fdg 19$ & \ion{O}{i} & $-155$--$-75$ & $0.16^{+0.04}_{-0.04}$--$0.39^{+0.21}_{-0.18}$ & 4 & $0.1\pm 0.1$ \\
		 & 3C~351 & $90 \fdg 08$ & $+36\fdg 38$ & \ion{O}{i} & $-128$ & $0.17^{+0.09}_{-0.08}$\tnote{g} & 5 & $0.08\pm 0.61$ \\
		 & & & & & & $0.12^{+0.02}_{-0.02}$--$0.21^{+0.04}_{-0.04}$ & 4 & \\
		 & Mrk~290 & $91\fdg 49$ & $+47\fdg 95$ & \ion{O}{i} & $-165$--$-75$ & $0.09^{+0.06}_{-0.03}$ & 4 & $0.02\pm 0.21$ \\
		 & Mrk~876 & $98\fdg 27$ & $+40\fdg 38$ & \ion{O}{i} & $-135$ & $0.20^{+0.08}_{-0.07}$ & 4 & $0$ \\
		 &  & & & & $-175$ & $0.26^{+0.13}_{-0.12}$ & & \\
		 & Mrk~817 & $100\fdg 30$ & $+53\fdg 48$ & \ion{O}{i} & $-140$--$-80$ & $0.29^{+0.11}_{-0.06}$ & 3, 4 & $0.2\pm 0.1$ \\
		 & & & & & $-148.6$, $-113.8$, $-86.0$ & $0.33^{+0.21}_{-0.17}$\tnote{g} & 6 & \\
		 & PG~1351$+$640 & $111\fdg 89$ & $+52\fdg 02$ & \ion{O}{i} & $-190$--$-95$ & $<0.38$ & 3, 4 & $0$ \\
		 & Mrk~279 & $115\fdg 04$ & $+46\fdg 86$ & \ion{O}{i} & $-200$--$-120$ & $0.11^{+0.04}_{-0.03}$ & 4 & $0.5\pm 0.6$ \\
		 & PG~1259$+$593 & $120\fdg 56$ & $+58\fdg 05$ & \ion{O}{i} & $-130$ & $0.093^{+0.125}_{-0.047}$ & 7 & $0.31\pm 0.04$ \\
		 & & & & \ion{O}{i} & $-155$--$-95$ & $0.10^{+0.05}_{-0.04}$ & 3, 4 \\
		 & & & & \ion{O}{i} & $-129.5$, $-112.5$ & $0.16^{+0.04}_{-0.06}$ & 8 & \\
		 & Mrk~205 & $125\fdg 45$ &  $+41\fdg 67$ & \ion{O}{i} & $-165$--$-125$ & $0.17^{+0.05}_{-0.05}$ & 4 & $0$ \\
		\hline
		Smith~Cloud & RX~J2043.1$+$0324 & $49 \fdg 72$ & $-22\fdg 88$ & \ion{S}{ii} & $+58.5$ & $0.72\pm 0.22\pm 0.25$\tnote{g} & 9 & $1.9\pm 0.1$ \\
		 & PG~2112$+$059 & $57 \fdg 04$ & $-28\fdg 01$ & & $+41.5$ & $0.81\pm 0.62\pm 0.28$\tnote{g} & & 0 \\
		 & RX~J2139.7$+$0246 & $58 \fdg 09$ & $-35\fdg 01$ & & $+53.7$ & $0.26\pm 0.12\pm 0.09$\tnote{g} & & 0 \\
		\hline
		Complex~WE & HD~156359 & $328\fdg 68$ & $-14\fdg 52$ & \ion{O}{i} & $+125.0$ & $2.3\pm 0.6$ & 10 & $\cdots$ \\
		\hline
		Complex~K & 3C~351 & $90 \fdg 08$ & $+36\fdg 38$ & \ion{O}{i} & $-82$ & $>0.30$\tnote{g} & 5 & $0.7\pm 0.6$ \\
		\hline
		Cloud~MIII & BD~$+$38$\degr$2182 & $182\fdg 16$ & $+62\fdg 21$ & \ion{O}{i} & $-85$ & $>0.27$ & 2, 11 & $0.6\pm 0.3$ \\
		\hline
		IV~Arch & HD~121800 & $113\fdg 01$ & $+49\fdg 76$ & \ion{S}{ii} & $-69$ & 0.78 & 2 & $0.84\pm 0.03$ \\
		 & & & & & $-37$ & 0.81& \\ 
		 & PG~1259+593 & $120\fdg 56$ & $+58\fdg 05$ & \ion{O}{i} & $-55$ & $0.98^{+1.21}_{-0.46}$ & 7 & $1.1\pm 0.1$ \\
		 & & & & & $-81.9$, $-54.5$, $-29.0$ & $0.79\pm 0.18$ & 8 \\
		 & PG~0953$+$414 & $179\fdg 79$ & $+51\fdg 71$ & \ion{S}{ii} & $-40$ & 1.1 & 2 & $1.4\pm 0.1$ \\
		 & HD~93521 & $183\fdg 14$ & $+62\fdg 15$ & \ion{S}{ii} & $-66.3$ & 2.1 & 2, 12 & $1.1\pm 0.2$ \\
		 & & & & & $-57.7$ & 0.78 & \\
		 & & & & & $-51.2$ & 1.2 & \\
		 & & & & & $-38.8$ & 0.74 & \\
		 & & & & & $\cdots$ & $1.0$\tnote{h} & \\
		\hline
		LLIV~Arch & PG~0804$+$761 & $138\fdg 28$ & $+31\fdg 03$ & \ion{O}{i} & $-55$ & 1.0 & 2 & $2.1\pm 0.1$ \\
		\hline
		PP~Arch & HD~215733 & $85\fdg 16$ & $-36\fdg 35$ & \ion{S}{ii} & $-93$ & 0.17 & 2, 13 & 0 \\
		 & & & & & $-56$ & 0.32 & \\
		 & & & & & $-43$ & 1.2 & \\
		\hline
		MS & NGC~7469 & $83\fdg 10$ & $-45\fdg 47$ & \ion{O}{i} & $-335$ & $0.10\pm 0.01 \pm 0.02$ & 14 & 0 \\
		 & NGC~7714 & $88\fdg 22$ & $-55\fdg 56$ & \ion{O}{i} & $-320$ & $0.06\pm 0.03$ & 15 & 0 \\
		 & HE~0056$-$3622 & $293\fdg 72$ & $-80\fdg 90$ & \ion{O}{i} & $+150$ & $0.09\pm 0.04$ & 15 & 0 \\
		 & Fairall~9 & $295\fdg 07$ & $-57\fdg 83$ & \ion{S}{ii} & $+125$--$+250$ & $0.28\pm 0.04^{+0.11}_{-0.14}$ & 16 & $0.06\pm 0.01$ \\
		 & & & & & $+143$--$+218$ & $0.50\pm 0.05$ & 17 \\
		 & RBS~144 & $299\fdg 48$ & $-65\fdg 84$ & \ion{S}{ii} & $+92$ & $0.07\pm 0.03$ & 15 & 0 \\
		\hline
	\end{tabular}
	\begin{tablenotes}
	\item[a] Name of the background object
	\item[b] Galactic coordinates of the background object
	\item[c] LSR velocity of the absorption line
	\item[d] Ion abundance relative to the solar value
	\item[e] References --- 1 \citet{1994A&A...282..709K}, 2 \citet{2001ApJS..136..463W}, 3 \citet{2003ApJ...585..336C}, 4 \citet{2007ApJ...657..271C}, 5 \citet{2003AJ....125.3122T}, 6 \citet{2023ApJ...946L..48F}, 7 \citet{2001ApJ...559..318R}, 8 \citet{2004ApJS..150..387S}, 9 \citet{2016ApJ...816L..11F}, 10 \citet{2023ApJ...944...65C}, 11 \citet{1993ApJ...416L..29D}, 12 \citet{1993ApJ...409..299S}, 13 \citet{1997ApJ...475..623F}, 14 \citet{2010ApJ...718.1046F}, 15 \citet{2013ApJ...772..110F}, 16 \citet{2000AJ....120.1830G}, 17 \citet{2013ApJ...772..111R}
	\item[f] Estimated dust-to-\ion{H}{i} ratio relative to the solar-neighbourhood value, toward the nearest neighbour of  the background object (this work)
	\item [g] With ionization correction
	\item [h] The average of the $-66.3$, $-57.7$, $-51.2$, and $-38.8$\,km\,s$^{-1}$ components, based on \ion{H}{i} column density
	\end{tablenotes}
	\end{threeparttable}
\end{table*}

\begin{figure}
\begin{center}
\includegraphics{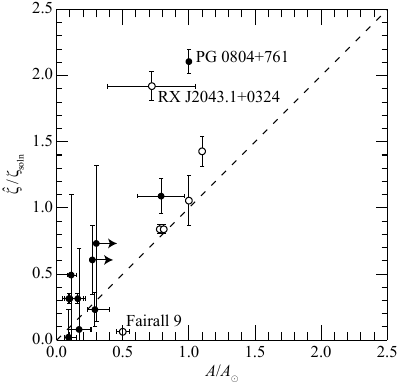}
\end{center}
\caption{
The correlation between the ion abundances $A/A_{\sun}$ obtained by absorption measurements (see Table \ref{tab:list_abs_measure}) and $\widehat{\zeta}/\zeta_{\mathrm{soln}}$ (this work).
The dashed line shows the line where both have the same value.
The filled and open symbols represent that the $A/A_{\sun}$ correspond to measurements from \ion{O}{i} and \ion{S}{ii} observations, respectively.
The arrows attached to the filled circles indicate that the $A/A_{\sun}$ values are the lower limit.
} \label{fig:abs_vs_zeta}
\end{figure}

The present work revealed the dust-to-\ion{H}{i} ratio distribution which covers a large fraction of the ISM, the IVCs, and the HVCs.
Dust only traces solid-phase heavy elements but is often, and also in the present study, used as a proxy for the gas-phase heavy elements assuming a nearly constant dust-to-metal (DTM) ratio.
Recent studies of nearby galaxies show that well-evolved galaxies with metallicities $12+\log(\mathrm{O}/\mathrm{H})\gtrsim 8.2$ \citep[cf. the solar value is 8.7, e.g.,][]{2009ARA&A..47..481A} have a more or less constant DTM ratio \citep[e.g.,][]{2019A&A...623A...5D}.
In the context of the low-velocity gas of the solar neighbourhood, these results support the assumption that the DTM ratio is constant and that the dust-to-\ion{H}{i} ratio is a good proxy for metallicity.

We demonstrate in Fig.~\ref{fig:abs_vs_zeta} that the present estimates are not far from the previous absorption-line measurements in Table~\ref{tab:list_abs_measure}, whereas the discrepancies toward PG~0804+761 (LL~IV~Arch), RX~J2043.1+0324 (the Smith Cloud) and Fairall~9 (the MS) suggest potential variations in the DTM ratio.
Based on the current limited observational data, it is hard to evaluate how generally the abovementioned assumption holds for IVCs and HVCs.

\subsection{Impact of the dust associated with the warm ionized medium on the estimates of the dust-to-\ion{H}{i} ratio}
\label{subsec:ionizedgas}
The diffuse warm ionized medium (WIM) outside of localized \ion{H}{ii} regions is another significant component of the ISM.
The absorption-line studies revealed that HVCs and IVCs are associated with the ionized components \citep[e.g.,][]{2003ApJS..146..165S,2009ApJ...699..754S,2011Sci...334..955L}.
The velocity-resolved high-sensitivity surveys of diffuse H~$\alpha$ emission using Wisconsin H~$\alpha$ Mapper (WHAM) showed that the neutral and ionized components trace each other well, though the detailed structure is not identical \citep[e.g.,][]{1998ApJ...504..773T,2001ApJ...556L..33H,2012ApJ...761..145B,2017ApJ...851..110B}.
The estimated mass of the associated ionized component is roughly comparable to that of the neutral counterpart.

The observational evidence for the WIM-related dust is reported \citep{1999ApJ...517..746H,2009ApJ...699.1374D,2019ApJ...887...89W}, but the extent to which it contributes to the FIR/submillimeter optical depth is poorly understood.
Previous works attempted to decompose the dust emission intensity into the WIM-related and \ion{H}{i}-related components.
\citet{1999A&A...344..322L,2000A&A...354..247L} claimed that they made a decomposition for the first time and that the FIR dust emissivity associated with WIM is close to the one with \ion{H}{i}.
On the contrary, \citet{2007ApJ...667...11O} reported that the WIM-related component is consistent with zero within the uncertainties.
\citet{2022ApJ...940..116C} reported that the FIR emission of the dust associated with a Reynolds layer of ionized hydrogen is below their detection limit.
These authors used regression models expressing dust emission intensity as a linear combination of $N_{\ion{H}{i}}$ and H~$\alpha$ emission intensity ($I_{\mathrm{H}\alpha}$).
We performed an alternative analysis by taking another approach.

In Section~\ref{sec:analyses}, we estimated the dust-to-\ion{H}{i} ratio $\zeta$ by making a least-squares fit to the regression model of equation (\ref{eqn:regression_model_libi}).
We refer to this as the ``without-WIM-terms model'' in this section and set up another ``with-WIM-terms model''
\begin{equation}
\begin{split}
\tau_{353}(l_{i}, b_{i}) = & \tau_{353, 0}(l_{i}, b_{i}) + {C_{0}}^{\alpha} \sum_{X} \left[ \zeta_{X}(l_{i}, b_{i}) W_{\ion{H}{i}, X}(l_{i}, b_{i}) \right]^{\alpha} \\
 & + \sum_{X}\tau_{353, \mathrm{H}^{+}, X}(l_{i}, b_{i}).
\end{split}
\end{equation}
under the assumption that there is some contribution to $\tau_{353}$ from the dust associated with the WIM ($\tau_{353, \mathrm{H}^{+}, X}$).
The dust optical depth is a function of column density, whereas $I_{\mathrm{H}\alpha}$ is the line-of-sight integral of the electron temperature $T_{\mathrm{e}}$, density $n_{\mathrm{e}}$ and ionized hydrogen density $n_{\mathrm{H}^{+}}$
\begin{equation}\label{eqn:halpha_intensity}
	I_\mathrm{H\alpha} = \frac{1}{2.75} \int \left(\frac{T_\mathrm{e}}{10^{4}\,\mathrm{K}}\right)^{-0.924} n_\mathrm{e}n_{\mathrm{H}^{+}}dl,
\end{equation}
where $dl$ is the line-of-sight path length over which the electrons are recombining.
The distribution of $T_\mathrm{e}$ and $n_\mathrm{e}\simeq n_{\mathrm{H}^{+}}$ in a line of sight is usually unknown.
Assuming that they are constant over the emitting region, we can approximate equation~(\ref{eqn:halpha_intensity}) as
\begin{equation} \label{eqn:emission_measure}
I_\mathrm{H\alpha} \propto {n_{\mathrm{H}^{+}}}^{2} L,
\end{equation}
and $N_{\mathrm{H}^{+}}$ is given as
\begin{equation} \label{eqn:wim_columndensity}
N_{\mathrm{H}^{+}} \simeq n_{\mathrm{H}^{+}}L \propto \left(L {I_\mathrm{H\alpha}}\right)^{\frac{1}{2}},
\end{equation}
where $L=\int dl$.
Then, we assume that the WIM term is proportional to ${I_\mathrm{H\alpha}}^{1/2}$.
The WHAM Sky Survey \citep{2003ApJS..149..405H,2010ASPC..438..388H} is the sole velocity-resolved all-sky H~$\alpha$ survey data presently available but provides us with no information on the high-velocity ($|V_{\mathrm{LSR}}|\gtrsim 100$\,km\,s$^{-1}$) WIM.
If $\tau_{353, \mathrm{H}^{+}, \mathrm{NHV}}$ and $\tau_{353, \mathrm{H}^{+}, \mathrm{PHV}}$ are tiny fraction, close to $\widehat{\tau}_{353, \ion{H}{i}, \mathrm{NHV}}$ and $\widehat{\tau}_{353, \ion{H}{i}, \mathrm{PHV}}$ (on the order of $10^{-8}$ or less), then approximating them to 0 would not significantly affect the results.
In summary, the WIM term is given as
\begin{equation}
\tau_{353, \mathrm{H}^{+}, X}(l_{i}, b_{i}) = \left\{
	\begin{array}{ll}
	0 & X=\mathrm{NHV}, \mathrm{PHV} \\
	\eta_{X}(l_{i}, b_{i}) \left[I_{\mathrm{H}\alpha, X}(l_{i}, b_{i})\right]^{\frac{1}{2}} & \mbox{otherwise}
	\end{array}
\right..
\end{equation}
We emphasize that the coefficients $\eta_{X}$ in the WIM terms are not dust-to-WIM ratios, unlike the coefficients $\zeta_{X}$ in the \ion{H}{i} terms.
Using $I_\mathrm{H\alpha}$ instead of $(I_\mathrm{H\alpha})^{1/2}$ almost reproduces the same results described below, indicating that the reality of the WIM terms in the model is not a very important issue in this discussion.

\begin{figure*}
\begin{center}
\includegraphics{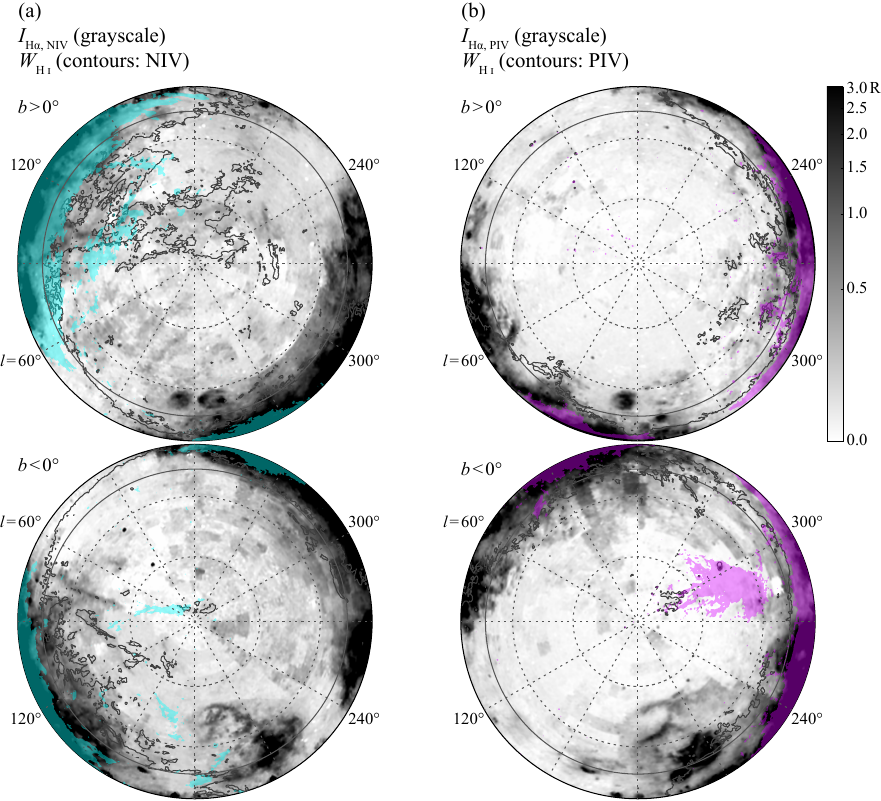}
\end{center}
\caption{
(a) The greyscale image shows the spatial distribution of H~$\alpha$ intensity $I_{\mathrm{H}\alpha}$ \citep{2003ApJS..149..405H,2010ASPC..438..388H} in the NIV velocity range shown in the same projection as Fig.~\ref{fig:HI_maps}.
The southern limit of the WHAM northern sky survey \citep{2003ApJS..149..405H}, $\delta_{\mathrm{J2000}}=-30\degr$, is presented by thick broken lines.
The contours show $W_{\ion{H}{i}, \mathrm{NIV}}=30$\,K\,km\,s$^{-1}$ in the same velocity range, and the cyan silhouette shows the NHV \ion{H}{i} components with $W_{\ion{H}{i}, \mathrm{PHV}}>10$\,K\,km\,s$^{-1}$.
The grey solid lines indicate $|b|=15\degr$.
The circular sectors bounded by thick solid lines present the two regions analyzed in Section~\ref{subsec:ionizedgas}.
(b) Same as (a) but for the PIV velocity range.
The contours show $W_{\ion{H}{i}, \mathrm{PIV}}=30$\,K\,km\,s$^{-1}$ and the magenta silhouette shows the PHV \ion{H}{i} components with $W_{\ion{H}{i}, \mathrm{PHV}}=10$\,K\,km\,s$^{-1}$.
} \label{fig:WHAM_maps}
\end{figure*}

\begin{figure}
\begin{center}
\includegraphics{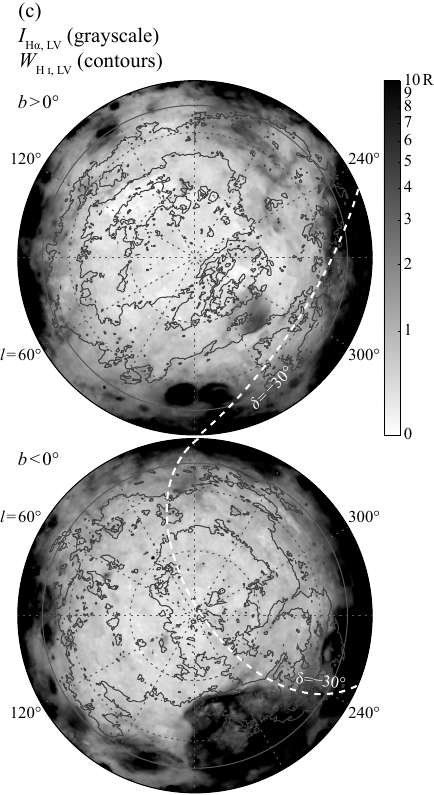}
\end{center}
\contcaption{
(c) Spatial distribution of $I_{\mathrm{H}\alpha}$ in the LV.
The contours show $W_{\ion{H}{i}, \mathrm{LV}}=30$, 100, and 300\,K\,km\,s$^{-1}$.
}
\end{figure}

\begin{figure*}
\begin{center}
\includegraphics{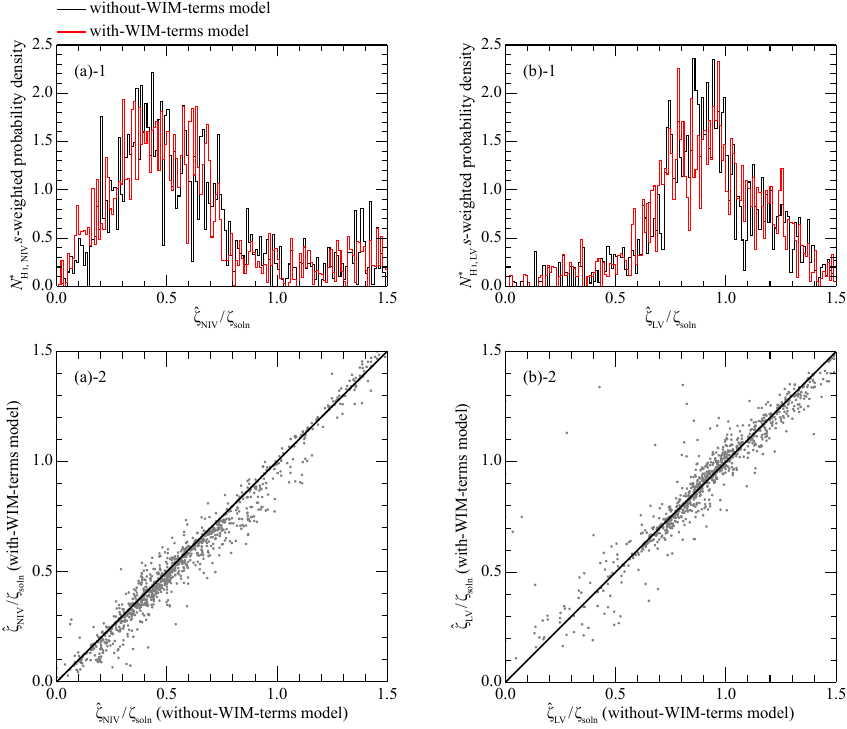}
\end{center}
\caption{
(a)-1 The black and red lines show the $N^{\ast}_{\ion{H}{i}, \mathrm{NIV}}s$-weighted probability density with respect to $\widehat{\zeta}_\mathrm{NIV}/\zeta_{\mathrm{soln}}$ obtained using the ``without-WIM-terms'' and ``with-WIM-terms'' models, respectively (see Section \ref{subsec:ionizedgas}).
(a)-2 Scatter plot showing a correlation between $\widehat{\zeta}_{\mathrm{NIV}}/\zeta_{\mathrm{soln}}$(without-WIM-terms model) and $\widehat{\zeta}_{\mathrm{NIV}}/\zeta_{\mathrm{soln}}$(with-WIM-terms model).
The solid line shows the line where both have the same value.
(b) Same as (a) but for $\widehat{\zeta}_{\mathrm{LV}}/\zeta_{\mathrm{soln}}$.
} \label{fig:with_vs_without_WIM_zeta}
\end{figure*}

\begin{figure}
\begin{center}
\includegraphics{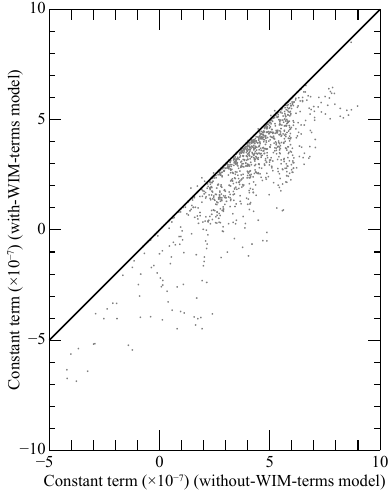}
\end{center}
\caption{
Scatter plot showing a correlation between constant term values of the ``without-WIM-terms model'' and those of the ``with-WIM-terms model''.
The solid line shows the line where both have the same value.
} \label{fig:with_vs_without_WIM_const}
\end{figure}

Fig.~\ref{fig:WHAM_maps}, the H~$\alpha$ intensity ($I_{\mathrm{H}\alpha}$) maps in the NIV, LV, and PIV velocity ranges, shows that the data suffer from zero-level offsets (typically $\sim 0.1$--0.2\,R, where $1\,\mathrm{R}=10^{6}/4\pi$\,photons\,cm$^{-2}$\,s$^{-1}$\,sr$^{-1}$) and block discontinuity, probably due to data-reduction issues.
We, therefore, used the data in $l=90\degr$--$150\degr$ and $b>45\degr$ in the following analysis, where the WHAM data are of acceptable quality (judged by the eye), and the optically-thick LV \ion{H}{I} occupies a small fraction of the area.
The WHAM-SS is undersampled with a 1\,deg beam at a 1\,deg spacing \citep[see][section~2.2]{2003ApJS..149..405H}, having a $\sim 3$--12 times lower resolution than the HI4PI and \textit{Planck} data, and we convolved the $W_{\ion{H}{i}}$ and $\tau_{353}$ data with the WHAM beam centred on each WHAM pointing.
Following \citet{2003ApJS..146..407F}, we approximated the WHAM beam to be a smoothed top-hat function
\begin{equation}
f(\theta)=\frac{1}{\exp\left[(\theta-\theta_{0})/\theta_\mathrm{s}\right]+1},
\end{equation}
where $\theta$ is the angular distance from the beam center, $\theta_{0}$ and $\theta_\mathrm{s}$ are set to 0.5 and 0.025\,deg, respectively.

Fig.~\ref{fig:with_vs_without_WIM_zeta}(a) compares the estimates of $\widehat{\zeta}_{\mathrm{NIV}}$ obtained using the ``without-WIM-terms model'' and those obtained using the ``with-WIM-terms model'', and Fig.~\ref{fig:with_vs_without_WIM_zeta}(b) compares $\widehat{\zeta}_{\mathrm{LV}}$.
The two models produce statistically similar results, although minor discrepancies exist in individual data points.
Fig.~\ref{fig:with_vs_without_WIM_const} shows a clear tendency that the constant term $\tau_{353, 0}$ in the "with-WIM-terms model" is somewhat smaller (note that this is only a comparison of the results from the two models and does not prove or measure $\tau_{353, \mathrm{H}^{+}, X}$, as it is uncertain how realistic the WIM terms in the model are).

In multiple regression analysis, generally, the constant term represents the portion of the dependent variable ($\tau_{353}$ in the present study) which is not significantly correlated with the independent variable ($W_{\ion{H}{i}, X}$ in the ``without-WIM-terms model'').
The contribution to $\tau_{353}$ from dust associated with WIM, if any, is included in the constant term in the ``without-WIM-terms model'' and has negligible impact on the estimates of $\zeta$ in the present study unless it is highly correlated with the \ion{H}{i} spatial distribution on a scale of $\sim 1$\,degree.

\subsection{The dust-to-gas ratio of the \ion{H}{i} components}
The derived dust-to-\ion{H}{i} ratio in the LV component and the four ROIs (Section~\ref{sec:analyses}) exhibits the following characteristics.
\begin{enumerate}
\item The ratio in the LV component has a distribution peaked at $\widehat{\zeta}_{\mathrm{LV}}/\zeta_{\mathrm{soln}}=1.0$ with a Gaussian shape having dispersion somewhat smaller but similar to the metallicities of G dwarfs in the local volume.
\item The NIV-GN region has a distribution with a significant fraction (25 per cent) of gas with a low ratio of $\widehat{\zeta}_{\mathrm{NIV}}/\zeta_{\mathrm{soln}}<0.5$.
In the NIV-GS region, PP~Arch and IVC~105$-$24 exhibit a low $\widehat{\zeta}_{\mathrm{NIV}}/\zeta_{\mathrm{soln}}\lesssim 0.3$.
\item The distribution in the NHV-GN peaked at $\widehat{\zeta}_{\mathrm{NHV}}/\zeta_{\mathrm{soln}}=0.1$--0.2, and more than 50 per cent of the fraction have ratios  $\widehat{\zeta}_{\mathrm{NHV}}/\zeta_{\mathrm{soln}}<0.3$.
\item The Magellanic Stream has the lowest distribution peaked at $\widehat{\zeta}/\zeta_{\mathrm{soln}}<0.1$ and more than 80 per cent of the fraction have ratios $\widehat{\zeta}/\zeta_{\mathrm{soln}}<0.2$. 
\end{enumerate}
We find that the LV, NHV-GN and the Magellanic Stream results are reasonably consistent with the previous absorption measurements, whereas the IVC results differ significantly from the previous results.

The conventional interpretation of the IVC is the Galactic fountain model, i.e., the stellar explosion in the plane feeds the high-metallicity gas to the Galactic halo to form the IVCs.
We argue that the present results require a model that the extragalactic low metallicity gas also supplies a significant fraction of the IVCs.
Otherwise, the significant fraction (20 per cent) of the gas with $\widehat{\zeta}_{\mathrm{NIV}}/\zeta_{\mathrm{soln}}<0.3$--0.5 cannot be explained.
Further, the interaction of the low metallicity IVCs with the high metallicity gas in the Galactic halo is an important mechanism to increase the metallicity in the IVCs.
\citet{2017ApJ...842..102G} and \citet{2017ApJ...837...82H} have recently discussed this possibility.
It is very likely that the IVCs are dynamically interacting with the Galactic halo gas, as shown in IVC~86$-$36 by \citetalias{2021PASJ...73S.117F}.
The interaction causes deceleration of the IVCs along with the accretion and mixing with the halo gas.
The interaction likely produces a trend that the metallicity, as well as the cloud mass, is increased along with the deceleration in the sense that the higher metallicity gas has a lower velocity.
The effect is approximately quantified as the mass increasing by a factor of two, accompanying a velocity decrease by a factor of two due to momentum conservation.
Concerning the metallicity, the interaction between an infalling cloud with a metallicity of 0.2 and the halo gas with a metallicity of 1.0 having the same mass will result in a metallicity of 0.6, which is closer to the halo metallicity than the infalling cloud, when the two gases are fully mixed.
The interaction has a strong impact on the observed metallicity of the IVCs.

\section{Concluding remarks}
\label{sec:conclusions}

We conducted a multiple-regression analysis combining 21\,cm \ion{H}{i} emission data with sub-mm dust optical depth $\tau_{353}$ to investigate the dust-to-\ion{H}{i} ratio in the \ion{H}{i} gas outside the Galactic plane.
We have resulted in a comprehensive dust-to-\ion{H}{i} ratio distribution for the IVCs, HVCs, Magellanic Stream, and the low-velocity component in the solar neighbourhood at an effective resolution of 47\,arcmin, which differs from the conventional optical absorption line measurements toward bright galaxies and stars that cover a tiny fraction of the gas.
The main conclusions are summarised below.
\begin{enumerate}
\item The present study allowed us to derive the dust-to-\ion{H}{i} ratio over a far greater portion of the \ion{H}{i} gas than the previous studies.
The major results include that the dust-to-\ion{H}{i} ratio of the IVCs (relative values to the solar-neighbourhood value) varies from $<0.2$ to $>1.5$ and that a significant fraction, 20 per cent, of the IVCs includes the dust-poor gas of $<0.5$ with a mode at 0.6.
In addition, it is shown that more than 50 per cent of the HVCs in Complex~C have ratios of $<0.3$ and that the Magellanic Stream has the lowest ratio with a mode at $\sim 0.1$.
\item We argue that a large fraction of the dust-poor IVC indicates that the infalling external \ion{H}{i} gas of low metallicity is significant in the IVCs.
A possible picture is that the low metallicity IVCs are falling onto the Galactic plane and are interacting with the high metallicity gas in the Galactic halo.
Such interaction is evidenced by the kinematic bridge features connecting the IVCs with the disc as observed, for instance, in IVC~86$-$36 in PP Arch, and will cause accretion of the halo gas onto the IVCs.
The accretion increases the metallicity and mass of the IVCs and decelerates their infall velocity, as shown by the theoretical work.
Therefore, the observed fraction of the low metallicity IVCs gives a secure lower limit for the fraction.
By the interaction, the IVC mass will be doubled accompanying a decrease of the infall velocity by a factor of 2 by the momentum conservation as well as the increase of the metallicity. This suggests that the fraction of the observed low metallicity IVCs should be doubled to 40 per cent prior to the onset of the collisional interaction in the halo.
Consequently, we suggest an overall picture that both the external infall and the Galactic fountain work to produce the IVCs, where the mass of the IVCs involved is comparable between the two mechanisms.
\item To pursue further the implications of the IVCs, we need more systematic studies of the \ion{H}{i} gas over the whole sky by focusing on the individual regions and the overall properties of the IVCs and HVCs.
The forthcoming new instruments, including SKA and ngVLA, are expected to provide important opportunities in these studies by covering the Local Group galaxies and nearby and more distant galaxies.
\end{enumerate}

\section*{Acknowledgements}

This work was supported by JSPS KAKENHI Grant Numbers JP15H05694 and JP21H00040.
This research made use of ds9, a tool for data visualisation supported by the Chandra X-ray Science Center (CXC) and the High Energy Astrophysics Science Archive Center (HEASARC) with support from the JWST Mission office at the Space Telescope Science Institute for 3D visualisation.
The valuable comments by Professors T.~Onishi, K.~Tachihara, A.~Mizuno and T.~Inoue helped to improve the content and readability of the paper.
We also thank the anonymous reviewers for their careful reading of our manuscript and their many valuable comments and suggestions.

\section*{Data Availability}

The HI4PI data underlying this article are available in VizieR at \url{http://cdsarc.u-strasbg.fr/viz-bin/qcat?J/A+A/594/A116}.
The \textit{Planck} data are in Planck Legacy Archive (PLA) at \url{https://pla.esac.esa.int/}.
The WHAM-SS data are available at \url{https://www.astro.wisc.edu/research/research-areas/galactic-astronomy/wham/wham-sky-survey/wham-ss-data-release/}.
The data products from this study will be shared on reasonable request to the corresponding author.



\bibliographystyle{mnras}
\bibliography{test,Planck_bib,statistics_bib,wakker01}



\appendix

\section{Consistency of the PR2 dust data with the PR1 data}
\label{sec:PR2_vs_PR1}

\begin{figure}
\begin{center}
\includegraphics{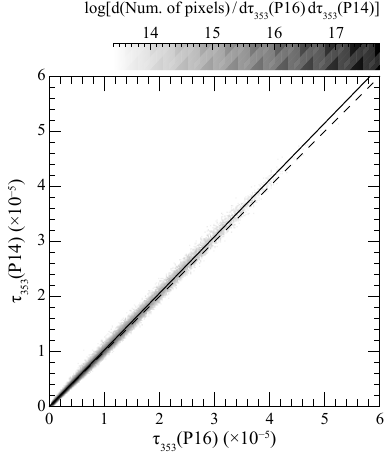}
\end{center}
\caption{
Density plot showing a correlation between $\tau_{353}\mbox{(P16)}$ and $\tau_{353}\mbox{(P14)}$.
The solid line shows the linear regression line (equation~(\ref{eqn:r2.0_vs_r1.2})) and the dashed line shows $\tau_{353}\mbox{(P14)}=\tau_{353}\mbox{(P16)}$.
}\label{fig:r2_vs_r1_tau353}
\end{figure}

We checked the consistency between \textit{Planck} PR2 GNILC dust data \citep[released version R2.01,][refer to it as the P16 data in this section]{planck2016-XLVIII} and PR1 data \citep[R1.20,][the P14 data]{planck2013-p06b}.
Here, we did not smooth the P16 data but smoothed the P14 data to have the same effective beam size as the original P16 data.
Then both were degraded to $N_\mathrm{side}=256$.
Here, we applied the masking criteria (a)--(c) in Section~\ref{subsec:masking}.

Fig.~\ref{fig:r2_vs_r1_tau353} shows the $\tau_{353}\mbox{(P16)}$-$\tau_{353}\mbox{(P14)}$ correlation.
The two datasets are highly correlated with a correlation coefficient of 0.999 and a linear regression by an OLS-bisector method \citep{1990ApJ...364..104I} is
\begin{equation}
\tau_{353}\mbox{(P14)} = (1.0297\pm 0.0002)\times \tau_{353}\mbox{(P16)}-(8.74\pm 0.07)\times 10^{-8} \label{eqn:r2.0_vs_r1.2}.
\end{equation}
We found a good consistency between $\tau_{353}$(P16) and $\tau_{353}$(P14).

\section{Geographically Weighted Regression Analysis}
\label{sec:GWR}

\subsection{Estimating the local coefficients}
\label{subsec:GWR_main}

\begin{figure}
\begin{center}
\includegraphics{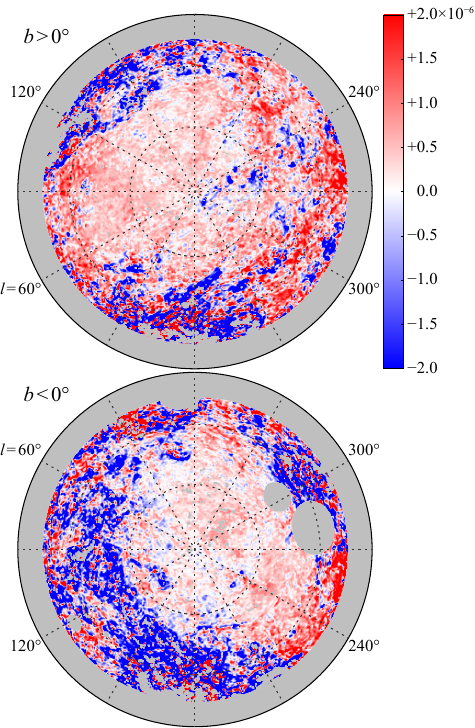}
\end{center}
\caption{
Spatial distribution of $\widehat{\tau}_{353, 0}$ shown in the same projection as Fig.~\ref{fig:HI_maps}.
}\label{fig:constant_maps}
\end{figure}

\begin{figure}
\begin{center}
\includegraphics{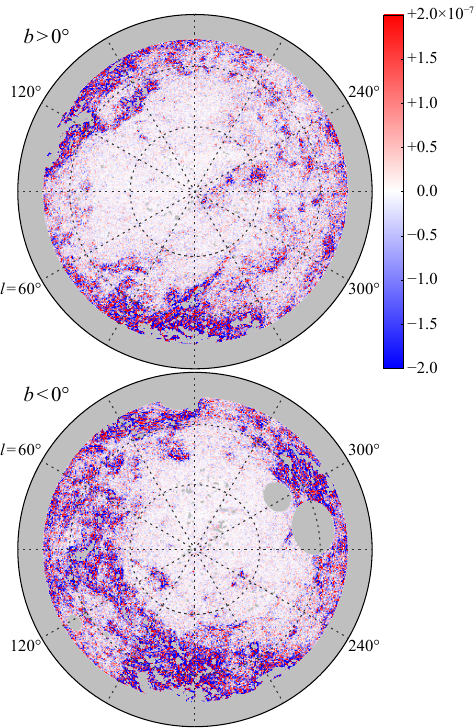}
\end{center}
\caption{
Spatial distribution of the residual from the regression shown in the same projection as Fig.~\ref{fig:HI_maps}.
}\label{fig:residual_maps}
\end{figure}

\begin{figure}
\begin{center}
\includegraphics{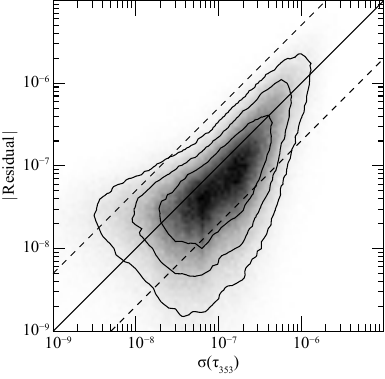}
\end{center}
\caption{
Density plot showing the correlation between $\sigma(\tau_{353})$ \citep{planck2016-XLVIII} and absolute values of the residuals from the regression.
The solid line shows the line where both have the same value, and the dashed lines show $|\mathrm{Residual}|=0.5\sigma{\tau_{353}}$ and $2\sigma{\tau_{353}}$.
The contours contain 50, 75 and 90 per cent of data points.
}\label{fig:residual_plot}
\end{figure}

Prior to the analysis, we performed a linearizing transformation on $W_{\ion{H}{i}, k}$ ($k=1, \cdots, m$; $m=5$ in the present study and corresponding to the five velocity ranges in Table~\ref{tab:velocity_components})
\begin{equation}
x_{k}(l_{i}, b_{i}) = \left\{
	\begin{array}{ll}
	{[W_{\ion{H}{i}, k}}(l_{i}, b_{i})]^{\alpha} & \mbox{if $W_{\ion{H}{i}, k}(l_{i}, b_{i}) > 5.5$\,K\,km\,s$^{-1}$}\\
	0 & \mbox{otherwise}
	\end{array}
\right.
\end{equation}
and replaced other variables,
\begin{eqnarray}
y(l_{i}, b_{i}) & = & {C_{0}}^{-\alpha}\tau_{353}(l_{i}, b_{i}) \\
a(l_{i}, b_{i}) & =& {C_{0}}^{-\alpha}\tau_{353, 0}(l_{i}, b_{i}) \\
\beta_{k}(l_{i}, b_{i}) & = & [\zeta_{k}(l_{i}, b_{i})]^{\alpha}.
\end{eqnarray}
Equation~(\ref{eqn:regression_model_libi}) was rewritten as
\begin{equation} \label{eqn:gwr_model}
y(l_{i}, b_{i}) = a(l_{i}, b_{i}) + \sum_{k=1}^{m}\left[\beta_{k}(l_{i}, b_{i})x_{k}(l_{i}, b_{i})\right] + \varepsilon(l_{i}, b_{i}),
\end{equation}
where $\varepsilon(l_{i}, b_{i})$ is the error term.
Equation~(\ref{eqn:gwr_model}) can be rewritten in a matrix form
\begin{equation}\label{eqn:gwr_model_matrix}
 \mathbfit{y} = \mathbfit{a} + (\mathbfss{B} \circ \mathbfss{X})\mathbfss{1}_{\bmath{m\times n}} + \boldsymbol{\varepsilon},
\end{equation}
where $\circ$ denotes the Hadamard product (also known as element-wise product) of the matrices,
\begin{equation}
\mathbfss{X} = \left[
	\begin{array}{ccc}
		x_{1}(l_{1}, b_{1}) & \cdots & x_{m}(l_{1}, b_{1}) \\
		\vdots & \ddots & \vdots \\
		x_{1}(l_{n}, b_{n}) & \cdots & x_{m}(l_{n}, b_{n})
	\end{array}
\right],
\end{equation}
is the matrix of the $n$ observations of $m$ independent variables,
\begin{equation}
\mathbfss{B} = \left[
	\begin{array}{ccc}
		\beta_{1}(l_{1}, b_{1}) & \cdots & \beta_{m}(l_{1}, b_{1}) \\
		\vdots & \ddots & \vdots \\
		\beta_{1}(l_{n}, b_{n}) & \cdots & \beta_{m}(l_{n}, b_{n})
	\end{array}
\right] = \left[
	\begin{array}{c}
		\boldsymbol{\beta}(l_{1}, b_{1})^\mathrm{T} \\
		\vdots \\
		\boldsymbol{\beta}(l_{n}, b_{n})^\mathrm{T}
	\end{array}
\right]
\end{equation}
is $n$ set of local coefficients, and $\mathbfss{1}_{\bmath{m\times n}}$ is an $m\times n$ all-ones matrix.
The $n$-element vectors $\mathbfit{y}$, $\mathbfit{a}$, and $\boldsymbol{\varepsilon}$ are dependent variables, local constant terms, and error terms.

The local regression coefficients at the $i$-th regression point are given by solving 
\begin{equation} \label{eqn:standard_GWR}
\widehat{\boldsymbol{\beta_{i}}}=\left(\mathbfss{X}_{\bmath{i}}^\mathrm{T}\mathbfss{W}_{\bmath{i}}\mathbfss{X}_{\bmath{i}}\right)^{-1}\mathbfss{X}_{\bmath{i}}^\mathrm{T}\mathbfss{W}_{\mathbfit{i}} \mathbfit{y}_{\bmath{i}},
\end{equation}
where 
\begin{equation}
\mathbfss{X}_{\bmath{i}} = \left(\mathbfss{I}_{\bmath{n}}-\mathbfss{W}_{\bmath{i}}\mathbfss{1}_{\bmath{n\times n}}\right)\mathbfss{X}
\end{equation}
is the matrix of the local-centred independent variables,
\begin{equation}
\mathbfit{y}_{\bmath{i}} = \left(\mathbfss{I}_{\bmath{n}}-\mathbfss{W}_{\bmath{i}}\mathbfss{1}_{\bmath{n\times n}}\right)\mathbfit{y},
\end{equation}
is the local-centred dependent variable, $\mathbfss{I}_{\bmath{n}}$ is the identity matrix of size $n$, $\mathbfss{1}_{\bmath{n\times n}}$ is an $n\times n$ all-ones matrix, $\boldsymbol{\beta_{i}}$ is a short-hand notation for $\boldsymbol{\beta}(l_{i}, b_{i})$, the superscript T indicates the matrix transpose, $\mathbfss{W}_{\bmath{i}}$ is a weighting matrix
\begin{equation}
\mathbfss{W}_{\bmath{i}} = \frac{1}{\sum_{j=1}^{n}w_{i}(l_{j}, b_{j})}\left[
	\begin{array}{ccc}
		w_{i}(l_{1}, b_{1}) & & \\
		 & \ddots & \\
		 & & w_{i}(l_{n}, b_{n})
	\end{array}
\right],
\end{equation}
and $w_{i}(l_{j}, b_{j})$ is given by equation~(\ref{eqn:weighting_function}).
As each component of $\boldsymbol{\beta_{i}}$ should be non-negative ($\boldsymbol{\beta_{i}}\geq 0$), equation~(\ref{eqn:standard_GWR}), therefore, must be reformulated by a NNLS problem
\begin{equation} \label{eqn:nnls_statement}
\mbox{Minimize}\ || \mathbfss{W}_{\bmath{i}}(\mathbfss{X}_{\bmath{i}}\boldsymbol{\beta_{i}}-\mathbfit{y}_{\bmath{i}})||\ \mbox{subject to}\ \boldsymbol{\beta_{i}}\geq 0,
\end{equation}
where $||\cdot ||$ denotes Euclidean norm (or also called $L^{2}$ norm).
A widely used algorithm for solving NNLS problems is the one by \citet{Lawson1995} briefly summarized as follows.
\begin{enumerate}
\renewcommand{\labelenumi}{\arabic{enumi}.}
\item Initialize set $P=\varnothing$ and $Q=\{1, \cdots, m\}$.
\item Let $\mathbfit{u}=(
	\begin{array}{ccc}
		u_{1} & \cdots & u_{m}
	\end{array}
)=\mathbfss{X}_{\bmath{i}}^\mathrm{T}\left(\mathbfit{y}_{\bmath{i}}-\mathbfss{X}_{\bmath{i}}\boldsymbol{\beta_{i}}\right)$.
\item If $Q=\varnothing$ or $\max\{u_{q}: q\in Q\}\leq 0$, the calculation is completed.
\item Find an index $k \in Q$ such that $u_{k}=\max\{u_{q}: q\in Q\}$, and then move the index $k$ from $Q$ to $P$.
\item Let $\mathbfss{P}_{\bmath{i}}$ denote the $n \times m$ matrix defined by 
	\begin{equation}
		\mbox{column $k$ of}\ \mathbfss{P}_{\bmath{i}}= \left\{\begin{array}{ll}
			\mbox{column $k$ of}\ \mathbfss{X}_{\bmath{i}} & \mbox{if}\ k \in P \\
			0 & \mbox{if}\ k \in Q.
		\end{array}\right.
	\end{equation}
\item Let $\mathbfit{z}$ be vector of same length as $\boldsymbol{\beta_{i}}$, and set 
	\begin{equation}
		\mathbfit{z} = (
		\begin{array}{ccc}
			z_{1} & \cdots & z_{m}
		\end{array}
	) = \left(\mathbfss{P}_{\bmath{i}}^\mathrm{T}\mathbfss{P}_{\bmath{i}}\right)^{-1}\mathbfss{P}_{\bmath{i}}^\mathrm{T}\mathbfit{y}.
	\end{equation}
\item Set $z_{q}=0$ for $q\in Q$.
\item If $\min\{z_{p}: p\in P\} > 0$, set $\boldsymbol{\beta_{i}}=\boldsymbol{z}$ and go to Step 2.
\item Let
	\begin{equation}
		\gamma=\min\left\{\frac{\beta_{p}}{\beta_{p}-z_{p}}: z_{p} \leq 0, p\in P \right\},
	\end{equation}
	where $\beta_{p}$ means the $p$-th element of $\boldsymbol{\beta_{i}}$.
\item Set $\boldsymbol{\beta_{i}}$ to $\boldsymbol{\beta_{i}}+\gamma(\mathbfit{z}-\boldsymbol{\beta_{i}})$.
\item Move from $P$ to $Q$ all indices $p\in P$ such that $\beta_{p} \leq 0$.
 Then go to Step 5.
\end{enumerate}
The obtained $\boldsymbol{\beta_{i}}$ satisfies $\beta_{p} > 0$ for $p\in P$, $\beta_{q} = 0$ for $q\in Q$, and is an estimator $\widehat{\boldsymbol{\beta_{i}}}$ for the least squares problem
\begin{equation}
	\mathbfss{P}_{\bmath{i}}\boldsymbol{\beta_{i}} = \mathbfit{y}_{\bmath{i}}.
\end{equation}
The estimator of the constant term is given by
\begin{equation}
\widehat{a}(l_{i}, b_{i}) = \mathbfss{1}_{\bmath{1\times n}}\mathbfss{W}_{\bmath{i}}\left[\mathbfit{y} - \mathbfss{X}\widehat{\boldsymbol{\beta_{i}}} \right],
\end{equation}
where $\mathbfss{1}_{\bmath{1\times n}}$ is an $n$-element all-ones row vector.

The estimated regression coefficients are transformed as following
\begin{eqnarray}
\widehat{\zeta}_{k}(l_{i}, b_{i}) & = & \left[\widehat{\beta}_{k}(l_{i}, b_{i})\right]^{1/\alpha} \\
\widehat{\tau}_{353, 0}(l_{i}, b_{i}) & = & {C_{0}}^{\alpha} \widehat{a}(l_{i}, b_{i}).
\end{eqnarray}
Figs.~\ref{fig:constant_maps} and Fig.~\ref{fig:residual_maps} shows the constant term $\widehat{\tau}_{353, 0}$ and the residual from the regression given by $\tau_{353}(l_{i}, b_{i})-\widehat{\tau}_{353}(l_{i}, b_{i})$, respectively.
Fig.~\ref{fig:residual_plot} shows that the residuals are comparable to $\sigma(\tau_{353})$, indicating the goodness-of-fit.
The residuals tend to be somewhat smaller, probably due to the smoothing applied to the \ion{H}{i} data (see Section \ref{subsec:preprocessing}).

\subsection{The standard error of the estimated local coefficients}
\label{subsec:GWR_stderr}

\begin{figure}
\begin{center}
\includegraphics{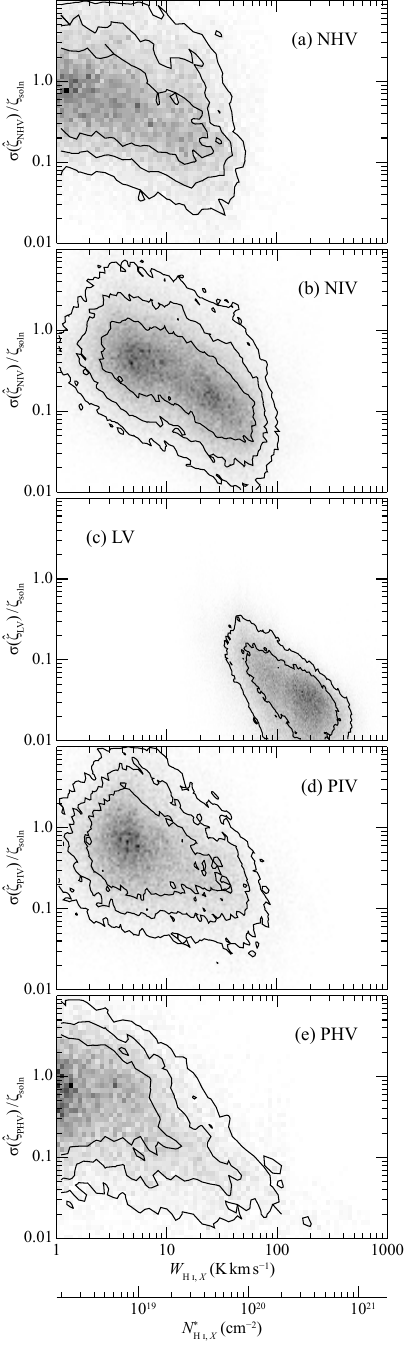}
\end{center}
\caption{
Density plot showing the correlation between $\sigma(\widehat{\zeta}_{X})/\zeta_{\mathrm{soln}}$ and $W_{\ion{H}{i}, X}$ in (a) NHV, (b) NIV, (c) LV, (d) PIV, and (e) PHV.
The contours contain 50, 75, and 90 per cent of data points.
} \label{fig:WHI_vs_SEzeta}
\end{figure}

Let 
\begin{equation}
\mathbfss{C}_{\bmath{i}} = \left(\mathbfss{P}_{\bmath{i}}^\mathrm{T}\mathbfss{W}_{\bmath{i}}\mathbfss{P}_{\bmath{i}}\right)^{-1}\mathbfss{P}_{\bmath{i}}^\mathrm{T}\mathbfss{W}_{\bmath{i}},
\end{equation}
where $\mathbfss{P}_{\bmath{i}}$ satisfies 
\begin{equation}
	\mbox{column $k$ of}\ \mathbfss{P}_{\bmath{i}}= \left\{\begin{array}{ll}
		\mbox{column $k$ of}\ \mathbfss{X}_{\bmath{i}} & \mbox{if the $k$-th element of $\widehat{\boldsymbol{\beta_{i}}} > 0$} \\
		0 & \mbox{if the $k$-th element of $\widehat{\boldsymbol{\beta_{i}}} = 0$}
	\end{array}\right..
\end{equation}
Then the estimated variance-covariance matrix of $\widehat{\boldsymbol{\beta}}(l_{i}, b_{i})$ is denoted by
\begin{equation}
\mathbfss{V}_{\mathbfit{i}} = \sigma^{2}\mathbfss{C}_{\bmath{i}}\mathbfss{C}_{\bmath{i}}^\mathrm{T},
\end{equation}
where $\sigma^{2}$ is the normalized residual sum of squares (RSS) from the local regression
\begin{equation}
\sigma^{2} = \frac{\left(\mathbfit{y}_{\bmath{i}}-\widehat{\mathbfit{y}_{\bmath{i}}} \right)^\mathrm{T} \mathbfss{W}_{\bmath{i}} \left(\mathbfit{y}_{\bmath{i}}-\widehat{\mathbfit{y}_{\bmath{i}}} \right) }{n-2\mathrm{tr}\mathbfss{S}+\mathrm{tr}(\mathbfss{S}^\mathrm{T}\mathbfss{S})} \sim \frac{ \left(\mathbfit{y}_{\bmath{i}}-\widehat{\mathbfit{y}_{\bmath{i}}} \right)^\mathrm{T} \mathbfss{W}_{\bmath{i}} \left(\mathbfit{y}_{\bmath{i}}-\widehat{\mathbfit{y}_{\bmath{i}}} \right) }{n-\mathrm{tr}\mathbfss{S}},
\end{equation}
where $\widehat{\mathbfit{y}_{\bmath{i}}} = \mathbfss{P}_{\bmath{i}}\widehat{\boldsymbol{\beta_{i}}}$.
The matrix $\mathbfss{S}$ is the hat matrix which maps $\widehat{\mathbfit{y}_{\bmath{i}}}$ on to $\mathbfit{y}_{\bmath{i}}$
\begin{equation}
\widehat{\mathbfit{y}_{\bmath{i}}} = \mathbfss{S}\mathbfit{y}_{\bmath{i}},
\end{equation}
and the $i$-th row of $\mathbfss{S}$ is given by 
\begin{equation}
\mathbfit{s}_{\bmath{i}} = (\mbox{row $i$ of } \mathbfss{P}_{\bmath{i}})\mathbfss{C}_{\bmath{i}}.
\end{equation}

The standard errors of the components of $\widehat{\boldsymbol{\beta}}(l_{i}, b_{i})$, $\sigma\left[\widehat{\beta}_{k}(l_{i}, b_{i})\right]$ are obtained from square roots of diagonal elements of $\mathbfss{V}_{\mathbfit{i}}$.
The estimated standard errors are transformed as
\begin{equation}
\sigma(\widehat{\zeta}_{k})(l_{i}, b_{i}) = \frac{1}{\alpha} \widehat{\beta}_{k} \left[\widehat{\beta}_{k}(l_{i}, b_{i})\right]^{1/\alpha-1} \sigma\left[\widehat{\beta}_{k}(l_{i}, b_{i})\right].
\end{equation}
Fig.~\ref{fig:WHI_vs_SEzeta} shows that $\sigma(\widehat{\zeta}_{X})$ is inversely proportional to $W_{\ion{H}{i}, X}$, this is because $\zeta_{X}$ are the gradient of the plane in the $\tau_{353}$-$W_{\ion{H}{i}, X}$ space.
The closer to the origin, the higher-$\zeta_{X}$- and lower-$\zeta_{X}$ planes become close.


\bsp	
\label{lastpage}
\end{document}